\documentclass[useAMS]{mn2e}
\usepackage{graphicx}
\usepackage{lscape}

\def\lesssim{\mathrel{\hbox{\rlap{\hbox{\lower4pt\hbox{$\sim$}}}\hbox{$<$}}}}

\def\gtrsim{\mathrel{\hbox{\rlap{\hbox{\lower4pt\hbox{$\sim$}}}\hbox{$>$}}}}

\title[Ensemble asteroseismology of  ZZ Ceti stars]
{Toward ensemble asteroseismology of ZZ Ceti stars with 
fully evolutionary models}

\author[A. D. Romero, A. H. C\'orsico, L. G. Althaus, et al.]
{A. D. Romero$^{1,2}$, 
A. H. C\'orsico$^{1,2}\thanks{E-mail: acorsico@fcaglp.unlp.edu.ar (AHC)}$, 
L. G. Althaus$^{1,2}$,
S. O. Kepler$^{3}$,
\newauthor 
B. G. Castanheira$^{3,4}$, and
M. M. Miller Bertolami$^{1,2}$\\
$^{1}$Facultad de Ciencias Astron\'omicas y Geof\'isicas, 
Universidad Nacional de La Plata, Argentina\\
$^{2}$CONICET\\
$^{3}$Departamento de Astronomia, Universidade Federal do 
Rio Grande do Sul, Av. Bento Goncalves 9500
Porto Alegre 91501-970, RS, Brazil\\
$^{4}$Institut f\"ur Astronomie, T\"urkenschanzstr. 17, A-1180 Wien, 
Austria}

\begin{document}

\date{}

\maketitle

\label{firstpage}

\begin{abstract}
ZZ  Ceti stars  form the  most numerous  group of  degenerate variable
stars. They are otherwise  normal DA (H-rich atmospheres) white dwarfs
that  exhibit  pulsations.  Here,  we  present an  asteroseismological
analysis  for 44 bright  ZZ Ceti  stars based  on a  new set  of fully
evolutionary DA white dwarf  models characterized by detailed chemical
profiles from  the centre to the  surface.  One of our  targets is the
archetypal  ZZ   Ceti  star  G117$-$B15A,  for  which   we  obtain  an
asteroseismological model with an  effective temperature and a surface
gravity   in   excellent  agreement   with   the  spectroscopy.    The
asteroseismological analysis of a set of 44 ZZ Ceti stars has
the potential to  characterize the global properties of  the class, in
particular  the thicknesses  of the  hydrogen envelope  and  the stellar
masses.  Our results support the belief that white dwarfs in the solar
neighbourhood harbor a broad range of hydrogen-layer thickness.
\end{abstract}


\section{Introduction}

Pulsating DA  (H-rich atmospheres) white dwarfs, commonly  known as ZZ
Ceti  or DAV  variable  stars,  comprise the  most  numerous class  of
compact pulsators.   They are  located in a  narrow and  probably pure
instability strip with effective  temperatures between $10\,500$ K and
$12\,500$ K (e.g.,  Winget \& Kepler 2008; Fontaine  \& Brassard 2008;
Althaus et al.  2010a).  ZZ  Ceti stars are characterized by multimode
photometric  variations  of  up  to  $0.30$ mag  caused  by  nonradial
$g$-mode pulsations of low degree  ($\ell \leq 2$) and periods between
70 and 1500 s.  The driving mechanism thought to excite the pulsations
is a  sort of combination  of the $\kappa-\gamma$ mechanism  acting in
the  hydrogen  partial  ionization   zone  (Dolez  \&  Vauclair  1981;
Dziembowski \& Koester 1981; Winget et al.  1982) and the ``convective
driving''  mechanism  proposed first  by  Brickhill  (1991) and  later
re-examined by Goldreich \& Wu (1999). The later mechanism is supposed
to  be dominant  once a  thick convection  zone has  developed  at the
stellar surface.

White-dwarf asteroseismology fully exploits the comparison between the
observed pulsation  periods in white  dwarfs and the  periods computed
for appropriate theoretical  models. It allows us to  infer details of
the origin,  internal structure and evolution of  white dwarfs (Winget
\& Kepler 2008; Fontaine \&  Brassard 2008; Althaus et al. 2010a).  In
particular, constraints on the   stellar mass, the thickness of the
outer envelopes,  the core chemical composition,  weak magnetic fields
and  slow rotation  rates can  be  inferred from  the observed  period
patterns of ZZ  Ceti stars.  In addition, asteroseismology  of ZZ Ceti
stars  is a valuable  tool for  studying axions  (Isern et  al.  1992;
C\'orsico  et al.   2001; Bischoff-Kim  et  al.  2008b;  Isern et  al.
2010, C\'orsico  et al.  2011), crystallization  (Montgomery \& Winget
1999; C\'orsico et  al.  2004, 2005; Metcalfe et  al.  2004; Kanaan et
al.   2005), and important  properties of  the outer  convection zones
(Montgomery  2005ab,  2007).  Finally,  the  temporal  changes in  the
observed  stable periods  allow  the measurement  of  the white  dwarf
evolutionary  timescale and  the detection  possible  planets orbiting
white dwarfs (Mullally et al.  2008).

Among  the  numerous  ZZ   Ceti  stars  currently  known  (148  stars;
Castanheira et al.  2010), in this  paper we will analyze 44 bright ZZ
Ceti stars which are listed in Table 1 of Fontaine \& Brassard (2008).
We  defer to a  future work  the study  of the  fainter ZZ  Ceti stars
discovered  within the  Sloan Digital  Sky Survey  (SDSS)  (Mukadam et
al. 2004; Mullally  et al.  2005; Kepler et  al. 2005b; Castanheira et
al. 2006,  2007, 2010).   The first target  star of  our seismological
survey is the most studied  member of the class, the paradigmatic star
G117$-$B15A.  This  star is an  otherwise typical DA white  dwarf, the
variability of which was discovered  by McGraw \& Robinson (1976) and,
since then, it has  been monitored continuously.  The surface gravity,
total  mass, and  effective temperature  of  this star  have been  the
subject  of  numerous  spectroscopic  determinations.   For  instance,
values of $\log g= 7.97 \pm  0.05$, $M_*= 0.59 \pm 0.03 \, M_{\odot}$,
and $T_{\rm eff}= 11\,630 \pm 200$ K, have been derived by Bergeron et
al.   (1995a, 2004) from  optical spectra.   Koester \&  Allard (2000)
have reported somewhat lower values for the gravity and mass, $\log g=
7.86  \pm  0.14$,  $M_*=  0.53\pm  0.07\,  M_{\odot}$,  and  a  higher
effective temperature, $T_{\rm  eff}= 11\,900 \pm 140$ K,  from HST UV
spectra.  G117$-$B15A  has oscillation periods  $\Pi$ (amplitudes $A$)
of 215.20 s  (17.36 mma), 270.46 s (6.14 mma) and  304.05 s (7.48 mma)
(Kepler et al.  1982) that correspond to genuine pulsation modes.  The
star also  shows the  harmonic of the  largest amplitude mode  and two
linear combinations.  Kepler et al. (2005a) used the rate of change of
the 215 s periodicity to show that the star has a C-O core.  The first
detailed  asteroseismological  study of  this  star  was presented  by
Bradley (1998).  This author obtained two different structures for the
star according  to the  assignation of the  radial order ($k$)  of the
modes exhibited  by the star.  If the  periods at 215, 271,  and 304 s
are associated with  $k= 1, 2, 3$, respectively,  this author obtained
an  asteroseismological model  with a  hydrogen envelope  mass $M_{\rm
  H}/M_*\sim 3 \times 10^{-7}$.  If,  instead, the periods have $k= 2,
3,  4$, the  asteroseismological  model was  characterized by  $M_{\rm
  H}/M_*\sim 1.5 \times 10^{-4}$.  Note that there are three orders of
magnitude of difference in the mass  of the H envelope between the two
possible   (and   nearly    equally   valid   within   their   models)
asteroseismological solutions.  A  similar degeneracy of seismological
solutions for G117$-$B15A  was also found by Benvenuto  et al.  (2002)
on  the basis  of independent  stellar and  pulsation  modeling.  More
recently,  Castanheira \&  Kepler  (2008) have  found a  seismological
solution  with $M_{\rm  H}/M_*  \sim 10^{-7}$  and  $k= 1,  2, 3$  and
another equally  valid solution with $M_{\rm H}/M_*  \sim 10^{-5}$ and
$k= 2,  3, 4$.  Finally, Bischoff-Kim  et al.  (2008a)  also found two
classes of  solutions, one characterized by ``thin''  H envelopes, and
other associated with ``thick''  H envelopes, although their ``thick''
envelope   solutions   ($M_{\rm  H}/M_*=   6   \times  10^{-7}$)   are
considerably thinner than those of the previous works.

Each of the mentioned  asteroseismological studies constitutes a clear
demonstration of the formidable capability of asteroseismology to shed
light  on the  internal  structure  of DA  white  dwarfs. However,  as
important  as  they  are,  all  of  these  studies  are  based  on  DA
white-dwarf  models that  lack a  fully consistent  assessment  of the
internal chemical  structure from the  core to the outer  layers.  For
instance, in the models of  Bradley (1998), although the C/He and He/H
chemical interfaces are more realistic than previous studies that used
the trace  element approximation (Tassoul  et al. 1990), the  core C-O
chemical profiles  have a (unrealistic) ramp-like shape.   In the case
of  Benvenuto et  al.  (2002),  the artificially-generated  models are
characterized   by   He/H  chemical   interfaces   resulting  from   a
time-dependent element diffusion treatment,  and the C-O core chemical
structure is extracted from the independent computations of Salaris et
al.  (1997).  So, there is no consistent coupling between the chemical
structure of the core and  the chemical stratification of the envelope
of the models.  On the other  hand, the recent works by Castanheira \&
Kepler  (2008, 2009)  are based  on DA  white-dwarf models  similar to
those  of  Bradley (1998),  with  a  parametrization  that mimics  the
results of time-dependent diffusion computations for the He/H chemical
interfaces,  but with  a  simplified treatment  of  the core  chemical
structure (50 \%  O and 50 \% C).  Finally,  the study of Bischoff-Kim
et  al.  (2008a)  employs DA  white-dwarf models  similar to  those of
Castanheira \& Kepler (2008, 2009),  but the envelope is stitched to a
core that  incorporates chemical profiles similar to  those of Salaris
et al. (1997).

Needless  to  say,  white-dwarf  stellar models  with  consistent  and
detailed chemical  profiles {\sl from  the centre to the  surface} are
needed to  correctly assess the  adiabatic pulsation periods  and also
the  mode-trapping properties  of  the DAVs,  the  crucial aspects  of
white-dwarf   asteroseismology  (Bradley   1996;   C\'orsico  et   al.
2002). In this regard, Althaus et al.  (2010b) (see also Renedo et al.
2010) have recently presented the first complete set of DA white-dwarf
models with consistent  chemical profiles for both the  core {\sl and}
the  envelope  for various  stellar  masses  appropriate for  detailed
asteroseismological fits  of ZZ  Ceti stars.  These  chemical profiles
are computed  from the full  and complete evolution of  the progenitor
stars from  the zero age main sequence,  through the thermally-pulsing
and mass-loss  phases on the  asymptotic giant branch (AGB),  and from
time-dependent  element diffusion  predictions during  the white-dwarf
stage.

In this paper, we  carry out the first asteroseismological application
of the  DA white-dwarf  models presented in  Althaus et  al.  (2010b).
Specifically, we perform a detailed asteroseismological study on 44 ZZ
Ceti stars that  includes the archetypal star G117$-$B15A,  by using a
grid of  new evolutionary models characterized  by consistent chemical
profiles and covering  a wide range of stellar  masses, thicknesses of
the    hydrogen   envelope    and    effective   temperatures.     The
asteroseismological  analysis of  such a  large set  of stars 
 is  a good starting  point for ensemble asteroseismology  of ZZ Ceti
stars (see Castanheira \& Kepler 2009). We also explore, in the frame
of standard  evolutionary calculations for  the formation of  DA white
dwarfs, to what extent the mass of the He-rich envelope ($M_{\rm He}$)
expected in DA white dwarfs  depends on the details of prior evolution
of progenitor  stars.  The  paper is organized  as follows.   In Sect.
\ref{numerical}, we  provide a  brief description of  the evolutionary
code, the  input physics adopted in  our calculations and  the grid of
models  employed.  There, we  also explore  the dependence  of $M_{\rm
  He}$ on the progenitor  evolution. In Sect.  \ref{astero-method}, we
describe our asteroseismological procedure. In Sect.  \ref{results} we
present  our results, starting  with the  asteroseismological analysis
for   G117$-$B15A    and   a   comparison    with   previous   results
(Sect. \ref{g117b15a}), and then by describing the results for the set
of   44   stars  (Sect.   \ref{zzcetis}).    We   conclude  in   Sect.
\ref{conclusions} by summarizing our findings.

\section{Numerical tools and models}
\label{numerical}

\subsection{Evolutionary code and input physics}

The present  asteroseismological study is  based on the full  DA white
dwarf evolutionary models of Althaus  et al.  (2010b) (see also Renedo
et al.  2010)  generated with the {\tt LPCODE}  evolutionary code.  To
our knowledge,  these models  are the first  complete set of  DA white
dwarfs models  characterized by consistent chemical  profiles for both
the core and envelope.  This feature renders these models particularly
suitable for asteroseismological studies of DA white dwarfs.

Here, we will briefly outline the most relevant characteristics of our
evolutionary  models  of  relevance  for their  pulsation  properties.
Further  details can  be  found in  Althaus  et al.   (2010b). In  our
computations,  the $^{12}$C($\alpha,\gamma)^{16}$O  reaction  rate, of
special  relevance for the  C-O stratification  of the  emerging white
dwarf, is  taken from  Angulo et al.   (1999).  Thus, our  white dwarf
models are characterized by  systematically lower central O abundances
than  the values predicted  by Salaris  et al.   (1997), who  used the
larger rate of Caughlan et al.   (1985). For example, for a $\sim 0.61
M_{\odot}$ white  dwarf, our  computations give $X_{\rm  O}\sim 0.73$,
about  $4 \%$ lower  than quoted  by Salaris  et al.   (1997) ($X_{\rm
  O}\sim  0.76$). Extra  mixing episodes  during core  He  burning, of
relevance  for  the  final  chemical stratification  of  white  dwarfs
(Straniero  et  al.   2003),   was  allowed  to  occur  following  the
prescription  of   Herwig  et   al.   (1997).   
Breathing pulses, which are convective runaways occurring towards  
the  end of  core  helium burning,  were
suppressed.   An  important  feature   of  our  computations  is  that
extra-mixing  episodes were  disregarded during  the thermally-pulsing
AGB phase, in line with theoretical and observational evidence (Herwig
et el. 2007, Lugaro et al.  2003, Salaris et al.  2009). This leads to
the inhibition  of the occurrence  of the third dredge-up  in low-mass
stars, and consequently, to  the gradual increase in the hydrogen-free
core (HFC) mass as evolution proceeds during this phase.  As a result,
the   initial-final   mass   relationship    by   the   end   of   the
thermally-pulsing AGB  is markedly different from  that resulting from
considering the mass of the  HFC right before the first thermal pulse.
This issue  is relevant  for the C-O  composition expected in  a white
dwarf.   Depending  on  the  white  dwarf  mass,  the  central  oxygen
abundance may be  underestimated by about 15 \% if  it is assumed that
the white dwarf  mass is the HFC mass by the  first thermal pulse (see
Althaus et al. 2010b).

We considered mass-loss episodes  during the core helium burning stage
and   on  the  red   giant  branch   following  Schr\"oder   \&  Cuntz
(2005). During  the AGB and  thermally-pulsing AGB phases,  we adopted
the maximum mass  loss rate between the prescription  of Schr\"oder \&
Cuntz (2005) and that of Vassiliadis \& Wood (1993).

\begin{table*}
\centering
\caption{The values of the stellar mass of our set of DA white dwarf models 
(upper row) and the mass of H corresponding to 
the different envelope thicknesses considered for each stellar mass.
The second row shows the maximum value of the thickness of the H envelope
for each stellar mass according to our evolutionary computations. }
\begin{tabular}{r|ccccccccccc}
\hline
\hline
$M_*/M_{\odot}$ &  0.5249 &0.5480 &0.5701 &0.5932 &0.6096 &0.6323 &0.6598 &0.7051 &0.7670 & 0.8373 &0.8779 \\
\hline
$\log(M_{\rm H}/M_*)$ & -3.62  & -3.74  & -3.82  & -3.93  & -4.02  & -4.12  & -4.25  & -4.45  & -4.70  & -5.00  & -5.07 \\
\hline
                     & -4.27  & -4.27  & -4.28  & -4.28  & -4.45  & -4.46  & -4.59  & -4.88  & -4.91  & -5.41  & -5.40 \\
                     & -4.85  & -4.85  & -4.84  & -4.85  & -4.85  & -4.86  & -4.87  & -5.36  & -5.37  & -6.36  & -6.39 \\
                     & -5.35  & -5.35  & -5.34  & -5.34  & -5.35  & -5.35  & -5.35  & -6.35  & -6.35  & -7.36  & -7.38 \\
                     & -6.33  & -6.35  & -6.33  & -6.33  & -6.34  & -6.34  & -6.35  & -7.35  & -7.34  & -8.34  & -8.37 \\
                     & -7.34  & -7.33  & -7.34  & -7.34  & -7.33  & -7.35  & -7.33  & -8.34  & -8.33  & -9.34  & -9.29 \\
                     & -8.33  & -8.33  & -8.31  & -8.33  & -8.33  & -8.33  & -8.33  & -9.34  & -9.33  &  ---   &   --- \\
                     & -9.25  & -9.22  & -9.33  & -9.33  & -9.25  & -9.34  & -9.33  &  ---   &  ---   &  ---   &   --- \\
\hline 
\end{tabular}
\label{table1}
\end{table*}

\begin{figure}
\begin{center}
\includegraphics[clip,width=8.3cm]{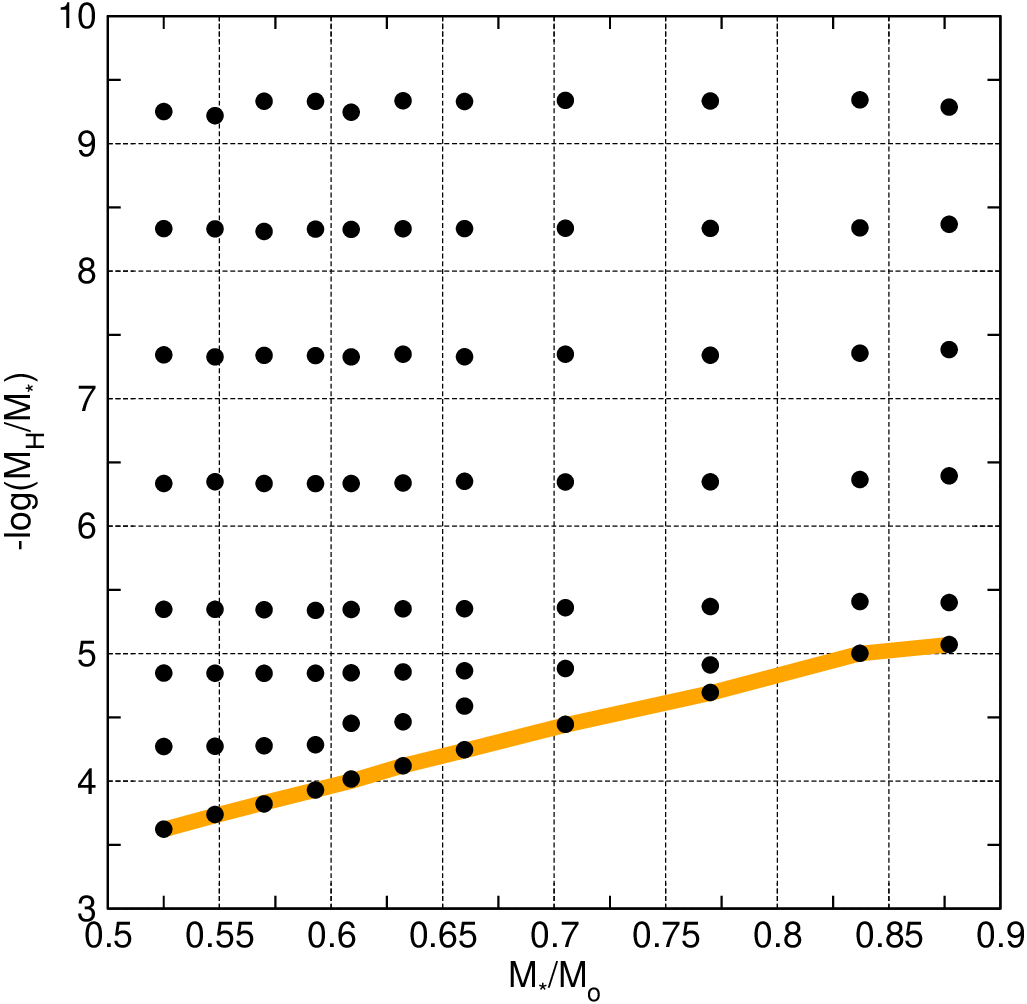}
\caption{The grid of DA  white dwarf evolutionary sequences considered
  in this study represented in  the plane $M_* - \log(M_{\rm H}/M_*)$.
  Each small circle corresponds to a sequence of DA white dwarf models
  with a given  stellar mass and thickness of  H envelope. The circles
  connected with a thick (orange) line correspond to the values of the
  maximum thickness of the H envelope as predicted by our evolutionary
  computations.   For each  sequence, we  have  pulsationally analysed
  about 200 stellar models covering the effective temperature range of
  $14\,000 - 9\,000$ K.}
\label{grid_mh}
\end{center}
\end{figure}

In  our evolutionary  computations,  we have  considered the  distinct
physical processes  that are responsible  for changes in  the chemical
abundance distribution  during white dwarf evolution.  This  is one of
the most important improvements of our computations in comparison with
previous asteroseismological works on DA white dwarfs.  In particular,
element diffusion  strongly modifies the  chemical composition profile
throughout their outer layers in the course of evolution.  As a result
of  diffusion processes, our  sequences develop  pure H  envelopes and
modifies  the  various  intershells  above  the  C-O  core.   We  have
considered  gravitational settling  as  well as  thermal and  chemical
diffusion --- but not radiative  levitation, which is relevant at high
effective temperatures for determining  the surface composition --- of
$^1$H, $^3$He, $^4$He, $^{12}$C,  $^{13}$C, $^{14}$N and $^{16}$O (see
Althaus et al.  2003 for  details).  The standard mixing length theory
for convection  --- with  the free parameter  $\alpha = 1.61$  --- has
been adopted.   Our treatment of time-dependent diffusion  is based on
the multicomponent gas treatment presented in Burgers (1969).  In {\tt
  LPCODE}, diffusion becomes operative  once the wind limit is reached
at high  effective temperatures (Unglaub \& Bues  2000).  In addition,
abundance changes  resulting from  residual nuclear burning  have been
taken  into account in  our simulations.   Finally, we  considered the
chemical rehomogenization  of the inner  carbon-oxygen profile induced
by  Rayleigh-Taylor  (RT)   instabilities  following  Salaris  et  al.
(1997).

An important  feature of our models  is the dependence  on the stellar
mass of  the outer layer  chemical stratification expected in  ZZ Ceti
stars.   Indeed,  for  the  more massive  models,  diffusion  strongly
modifies  the  chemical  abundance  distribution,  eroding  the  thick
intershell  region below  the  He  buffer by  the  time evolution  has
reached the  domain of the ZZ  Ceti instability strip  (see Althaus et
al. 2010b).  This is in contrast with the situation encountered in our
less  massive  models  ($M_*  \lesssim  0.63  M_{\odot}$),  where  the
intershell region is not removed by diffusion. This is because element
diffusion  is  less  efficient   in  less  massive  models  (with  the
subsequent longer diffusion timescale) and also because the intershell
is thicker  in these models.  Regarding  white dwarf asteroseismology,
these are  not minor  issues, since the  presence of  a double-layered
structure  in  the  helium-rich  layers  is  expected  to  affect  the
theoretical $g$-mode  period spectra  of ZZ Ceti  stars.  It  is clear
that white dwarf  evolution computed in a consistent  way with element
diffusion  as  considered  in  this  study  is  required  for  precise
asteroseismology.

\subsection{About the He content of a DA white dwarf star}
\label{hecontent}

\begin{figure}
\begin{center}
\includegraphics[clip,width=240pt]{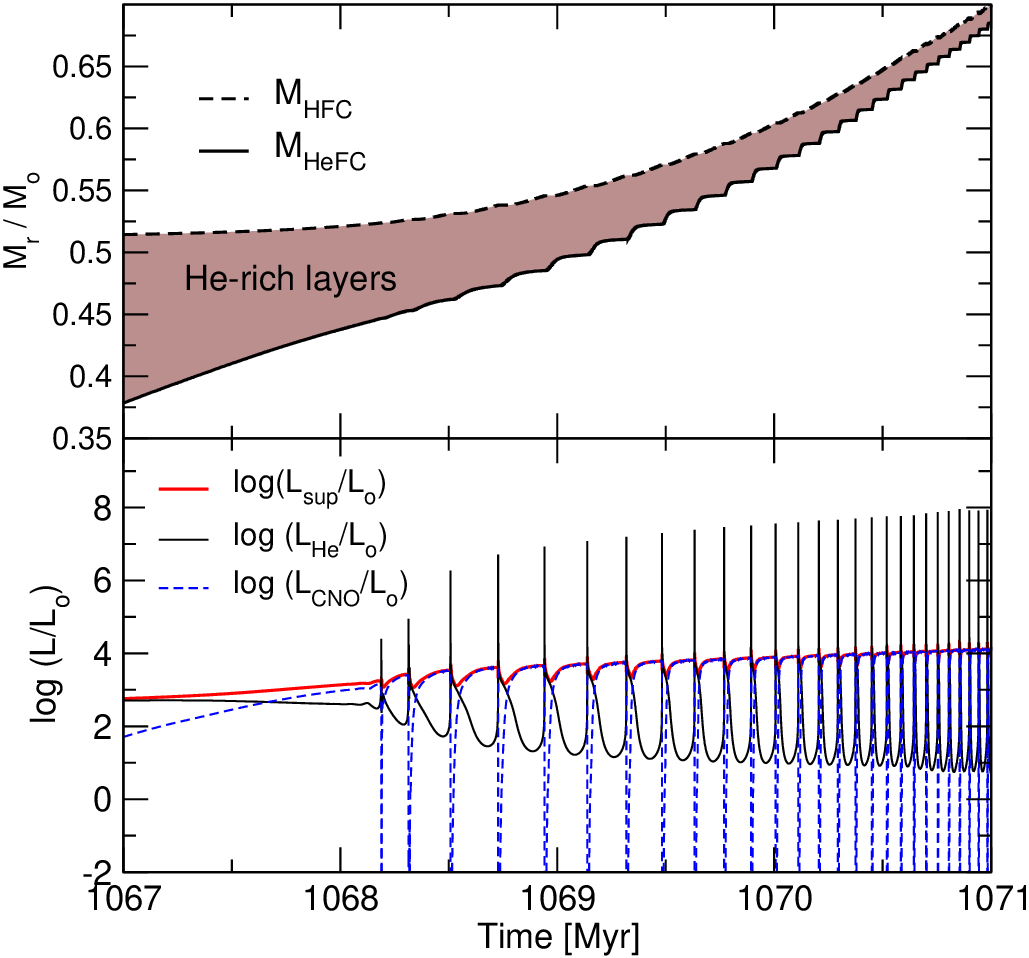}
\caption{Upper panel: change  in the He content in  the region limited
  by   the  boundaries   of  the   HeFC   and  the   HFC  during   the
  thermally-pulsing AGB phase. Lower  panel: the temporal evolution of
  surface luminosity and H- and He-burning luminosities in solar units
  for   our    initially   $2   M_{\odot}$-star    star   during   the
  thermally-pulsing AGB phase.}
\label{max-he}
\end{center}
\end{figure}

In this  section we show that,  in the frame  of standard evolutionary
computations for the  formation of DA white dwarfs,  the He content of
these stars cannot be substantially smaller than that predicted by our
calculations.   To  do  this,  we   compute  the  evolution  of  a  $2
M_{\odot}$-star from the ZAMS until the thermally-pulsing phase on the
AGB.  The  only way  we envisage  in which the  star may  experience a
substantial decrease  in its  content of He  is by undergoing  a large
number of thermal  pulses.  In order for the  model star to experience
the largest possible  number of thermal pulses, and  thus, the content
of He decreases as much as  possible, we switched off mass loss during
this stage  in our evolutionary  code. The results of  this experiment
are depicted in Fig. \ref{max-he}, in  which we show the He content in
the region  limited by the boundaries  of the He-free  core (HeFC) and
the HFC  in terms  of time during  the thermally pulsing  phase (upper
panel),  and  the  surface   luminosity  and  the  H-  and  He-burning
luminosities for each  pulse in that phase (lower  panel).  We stopped
the experiment  when the object  experienced about 30  thermal pulses,
which is enough for our purposes.  We found that the He content of the
object  decreased  from $M_{\rm  He}/M_{\odot}=  3.34 \times  10^{-2}$
(before the first thermal  pulse) to $M_{\rm He}/M_{\odot}= 8.6 \times
10^{-3}$ (before the thirtieth  thermal pulse). Thus, the decrease (in
solar  masses)  of  the  He  content   of  the  HFC  is  of  a  factor
$3.89$. However, it should be kept  in mind that this reduction is due
mainly to  the increase of  the mass of  the future white  dwarf, that
grows  from $M_{\rm  fWD}= 0.523  M_{\odot}$ to  $M_{\rm  fWD}= 0.7114
M_{\odot}$ between the thermal pulses 1 to 30.

Our experiment shows  that the He mass left in a  DA white dwarf could
be as much as a {\it  factor} $3-4$ lower than the values predicted by
standard  evolutionary computations,  but not  2 or  3 {\it  orders of
  magnitude} lower,  which would be necessary for  $g$-mode periods to
be substantially affected.  We conclude  that we can safety ignore the
variation of  $M_{\rm He}$ in  our asteroseismological analysis  of ZZ
Ceti stars. 

\subsection{The model grid}
\label{grid}

The DA  white dwarf models  employed in this  study are the  result of
full   evolutionary  calculations  of   progenitor  stars   for  
solar-like metallicity   ($Z=   0.01$).   The   complete   evolution  
of   eleven
evolutionary  sequences with initial  stellar mass  in the  range $1-5
M_{\odot}$   has   been   computed   from   the   ZAMS   through   the
thermally-pulsing and mass-loss  phases on the AGB and  finally to the
domain of planetary nebulae. The values of the stellar mass of our set
of models is  shown in the upper row of  Table \ref{table1}. The range
of stellar  mass covered by our computations  comfortably accounts for
the stellar mass of most of the observed pulsating DA white dwarfs.

Our asteroseismological approach  basically consists in the employment
of detailed white dwarf models characterized by very accurate physical
ingredients.   These models  are  obtained by  computing the  complete
evolution of  the progenitor stars. We have  applied successfully this
approach to the hot DOVs or  GW Vir stars (see C\'orsico et al. 2007a,
2007b, 2008, 2009).  Since  the final chemical stratification of white
dwarfs is fixed  in prior stages of their  evolution, the evolutionary
history of progenitor stars is  of utmost importance in the context of
white dwarf asteroseismology.  Our asteroseismological approach, while
being  physically  sounding,  is  by  far  much  more  computationally
demanding than  other approaches in which simplified  models are used.
As  a result,  our approach  severely  limits the  exploration of  the
parameter  space of  the models.   Indeed, for  the case  of  DA white
dwarfs, we  have only two  parameters which we  are able to vary  in a
consistent way: the stellar mass ($M_*$) and the effective temperature
($T_{\rm eff}$).   Instead, the thickness  of the H  envelope ($M_{\rm
  H}$),  the  content of  He  ($M_{\rm He}$),  the  shape  of the  C-O
chemical structure  at the core (including the  precise proportions of
central O and C), and the thickness of the chemical transition regions
are fixed by the evolutionary history of progenitor stars.  Therefore,
to push on the limits of our asteroseismological exploration, it would
be desirable to change  some additional parameters besides the stellar
mass and effective temperature of our DA models. In this work, we have
chosen  to  vary the  thickness  of the  H  envelope,  because of  the
uncertainties in the mass  loss rates.  According to full evolutionary
computations  (Althaus et  al.  2010b),  the maximum  H  envelope mass
expected in a white dwarf depends  on the stellar mass and ranges from
$M_{\rm H}/M_*  \sim 2.4 \times 10^{-4}$ (for  $M_*= 0.525 M_{\odot}$)
to $8.5 \times 10^{-6}$ (for $M_*= 0.878 M_{\odot}$)(see the first row
of Table  \ref{table1}).  Our decision for changing  this parameter is
due to several  reasons: first, there are compelling theoretical 
reasons to believe that the H-content of DA white dwarfs might depend 
on the details of their previous evolution. On the contrary, the He 
content or the inner C-O chemical profiles are not expected to
vary significantly due to the details of the previous evolutionary 
history (with the exception of a possible merger origin for the white 
dwarfs). Indeed, the total H  content remaining in some
DA white dwarfs  could be several orders of  magnitude lower than that
predicted  by  our standard  treatment  of  progenitor evolution.  For
instance, Althaus et  al. (2005b) have found that  $M_{\rm H}$ becomes
considerably  reduced if  the  progenitor experiences  a late  thermal
pulse   episode   (LTP)  shortly   after   the   departure  from   the
thermally-pulsing  AGB phase.   In  this sense,  Tremblay \&  Bergeron
(2008) show that the increase in  the ratio of He-rich to H-rich white
dwarfs can  be understood  on the  basis that a  fraction of  DA white
dwarfs above  $T_{\rm eff} \approx  10\,000$ K are characterized  by a
broad range of H-layer thickness.  Second, the precise location of the
He/H transition region (and the value of $M_{\rm H}$) strongly affects
the structure  of the  adiabatic period spectrum  in a DA  white dwarf
(Bradley 1996). Finally, $M_{\rm  H}$ is the structural parameter that
can be  more easily modified  in our models without  removing relevant
features predicted by the complete progenitor evolution.

In  order to  get different  thicknesses of  the H  envelope,  we have
followed a simple  recipe. For each sequence characterized  by a given
stellar mass  and a thick  value of $M_{\rm  H}$, as predicted  by the
full computation of the pre-white dwarf evolution (second row of Table
\ref{table1}),  we have  simply replaced  $^{1}$H by  $^{4}$He  at the
basis  of  the  H envelope.   This  is  done  at very  high  effective
temperatures ($\gtrsim 70\, 000$ K), in such a way that the unphysical
transitory  effects associated  to this  procedure end  much  long 
 before the
models reach the stage of pulsating DA white dwarfs. After our {\sl ad
  hoc} procedure to  change the thickness of the  H envelope, we allow
time-dependent element diffusion to operate while the models cool down
until  they reach  the effective  temperatures characterizing  the DAV
instability strip. Diffusion leads to very smooth chemical profiles at
the  He/H chemical transition  regions.  The  resulting values  of the
H content for the different envelopes are shown in Table \ref{table1},  
and a graphical representation of the basic  grid of models employed in this
work  is  displayed  in  Fig.   \ref{grid_mh}.  In  this  figure,  the
canonical  values of $M_{\rm  H}$ predicted  by stellar  evolution are
connected  with   a  thick   (orange)  line.  Obviously,   beyond  the
availability of the models of this coarse grid, we have the capability
to  generate additional  DA  white dwarf  evolutionary sequences  with
arbitrary  values of $M_{\rm  H}$ for  each stellar  mass in  order to
refine the model grid.

\subsection{Pulsation computations}

We carried  out the adiabatic  pulsation computations required  by the
present  asteroseismological  analysis   by  employing  the  nonradial
pulsation code described in C\'orsico \& Althaus (2006).  Briefly, the
code, which is coupled to the {\tt LPCODE} evolutionary code, is based
on  the   general  Newton-Raphson   technique  and  solves   the  full
fourth-order set  of equations governing  linear, adiabatic, nonradial
stellar   pulsations  following   the  dimensionless   formulation  of
Dziembowski (1971).   The prescription used  to assess the run  of the
Brunt-V\"ais\"al\"a   frequency  ($N$)   is  the   so-called  ``Ledoux
modified''  treatment   (see  Tassoul  et   al.   1990)  appropriately
generalized  to include the  effects of  having three  nuclear species
varying in abundance.

\begin{figure*}
\begin{center}
\includegraphics[clip,width=450pt]{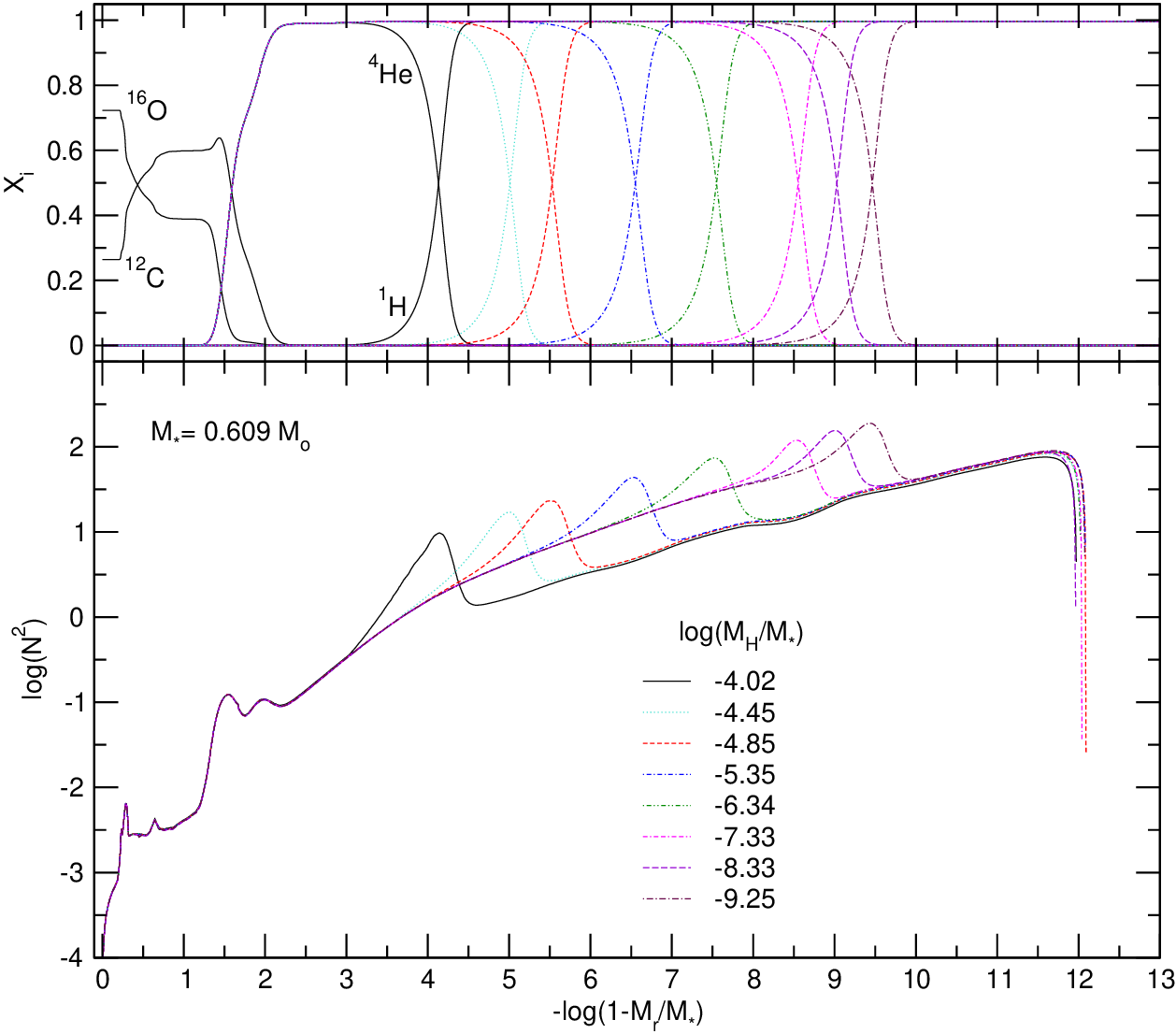}
\caption{Upper panel: the internal chemical profiles of DA white dwarf
  models with  $M_*= 0.609 M_{\odot}$,  $T_{\rm eff} \sim  12\,000$ K,
  and different thicknesses  of the H envelope. Only  the main nuclear
  species are depicted.  Lower panel:  the run of the logarithm of the
  squared  Brunt-V\"ais\"al\"a  frequency for  each  model.  Note  the
  correspondence between the chemical transition regions (upper panel)
  and the  resulting features in the shape  of the Brunt-V\"ais\"al\"a
  frequency.  For details, see the text.}
\label{chemical_bruntv}
\end{center}
\end{figure*}

\begin{table}
\caption{Atmospheric parameters and spectroscopic masses for the 
sample of ZZ Ceti stars analysed in this paper.} 
\centering
\scalebox{0.9}[0.9]{
\hspace{-7mm}
\begin{tabular}{lcccc}
\hline\hline
${\rm Star}$  &   $T_{\rm eff}$ [K] &    $\log g$    & $M_*/M_{\odot}$   &  Ref.\\
\hline\hline
G226$-$29       & $12\,460 \pm 200$  & $8.28\pm 0.05$  & $0.771\pm 0.032$ &    3 \\
HS 1531$+$7436  & $12\,350 \pm 181$  & $8.17\pm 0.048$ & $0.704\pm 0.029$ &    1 \\
G185$-$32       & $12\,130 \pm 200$  & $8.05\pm 0.05$  & $0.634\pm 0.028$ &    3 \\
L19$-$2         & $12\,100 \pm 200$  & $8.21\pm 0.05$  & $0.726\pm 0.033$ &    3 \\
G132$-$12       & $12\,080 \pm 200$  & $7.94\pm 0.05$  & $0.575\pm 0.026$ &    4 \\
EC 11507$-$1519 & $12\,030 \pm 200$  & $7.98\pm 0.05$  & $0.596\pm 0.026$ &    4 \\
PG 1541$+$650   & $12\,000 \pm 70 $  & $7.79\pm 0.04$  & $0.502\pm 0.023^{*}$ &6 \\
R548            & $11\,990 \pm 200$  & $7.97\pm 0.05$  & $0.590\pm 0.026$ &    3 \\
GD 165          & $11\,980 \pm 200$  & $8.06\pm 0.05$  & $0.639\pm 0.029$ &    3 \\
GD 66	        & $11\,980 \pm 200$  & $8.05\pm 0.05$  & $0.634\pm 0.028$ &    3 \\
G207$-$9        & $11\,950 \pm 200$  & $8.35\pm 0.05$  & $0.812\pm 0.033$ &    3 \\
EC 14012$-$1446 & $11\,900 \pm 200$  & $8.16\pm 0.05$  & $0.696\pm 0.031$ &    3 \\
KUV 11370$+$4222 &$11\,890 \pm 200$  & $8.06\pm 0.05$  & $0.639\pm 0.028$ &    3 \\
G238$-$53         & $11\,890 \pm 200$  & $7.91\pm 0.05$  & $0.559\pm 0.025$ &    3 \\
GD 99           & $11\,820 \pm 200$  & $8.08\pm 0.05$  & $0.650\pm 0.028$ &    3 \\
G29$-$38        & $11\,820 \pm 200$  & $8.14\pm 0.05$  & $0.684\pm 0.030$ &    3 \\
LP 133$-$144    & $11\,800 \pm 200$  & $7.87\pm 0.05$  & $0.539\pm 0.025$ &    3 \\
HS 1249$+$0426  & $11\,770 \pm 181$  & $7.92\pm 0.048$ & $0.564\pm 0.024$ &    1 \\
MCT 2148$-$2911 & $11\,740 \pm 200$  & $7.82\pm 0.05$  & $0.515\pm 0.023^{*}$ &5 \\
GD 385	        & $11\,710 \pm 200$  & $8.04\pm 0.05$  & $0.627\pm 0.028$ &    3 \\
GD 244          & $11\,680 \pm 200$  & $8.08\pm 0.05$  & $0.650\pm 0.028$ &    2 \\
HS 0507$+$0434B & $11\,630 \pm 200$  & $8.17\pm 0.05$  & $0.702\pm 0.030$ &    3 \\
G117$-$B15A     & $11\,630 \pm 200$  & $7.97\pm 0.05$  & $0.589\pm 0.026$ &    3 \\
EC 23487$-$2424 & $11\,520 \pm 200$  & $8.10\pm 0.05$  & $0.661\pm 0.028$ &    3 \\
MCT 0145$-$2211 & $11\,500 \pm 200$  & $8.14\pm 0.05$  & $0.684\pm 0.030$ &    3 \\
KUV 08368$+$4026 &$11\,490 \pm 200$  & $8.05\pm 0.05$  & $0.633\pm 0.028$ &    3 \\
PG 2303$+$243   & $11\,480 \pm 200$  & $8.09\pm 0.05$  & $0.655\pm 0.028$ &    3 \\
BPM 31594       & $11\,450 \pm 200$  & $8.11\pm 0.05$  & $0.666\pm 0.029$ &    3 \\
HLTau$-$76      & $11\,450 \pm 200$  & $7.89\pm 0.05$  & $0.548\pm 0.025$ &    3 \\
G255$-$2        & $11\,440 \pm 200$  & $8.17\pm 0.05$  & $0.702\pm 0.030$ &    3 \\
HE 1429$-$037   & $11\,434 \pm 36 $  & $7.82\pm 0.02$  & $0.514\pm 0.010^{*}$ &7 \\
G191$-$16       & $11\,420 \pm 200$  & $8.05\pm 0.05$  & $0.632\pm 0.028$ &    3 \\
HE 1258$+$0123  & $11\,400 \pm 200$  & $8.04\pm 0.05$  & $0.627\pm 0.029$ &    3 \\
G232$-$38       & $11\,350 \pm 200$  & $8.01\pm 0.05$  & $0.610\pm 0.027$ &    4 \\
KUV 02464$+$3239 & $11\,290 \pm 200$  & $8.08\pm 0.05$  & $0.648\pm 0.028$ &    2 \\
HS 1625$+$1231  & $11\,270 \pm 181$  & $8.06\pm 0.048$ & $0.638\pm 0.027$ &    1 \\
BPM 30551       & $11\,260 \pm 200$  & $8.23\pm 0.05$  & $0.737\pm 0.032$ &    3 \\
HS 1824$-$6000  & $11\,192 \pm 181$  & $7.65\pm 0.048$ & $0.427\pm 0.030^{*}$ &1 \\
G38$-$29        & $11\,180 \pm 200$  & $7.91\pm 0.05$  & $0.557\pm 0.025$ &    3 \\
GD 154          & $11\,180 \pm 200$  & $8.15\pm 0.05$  & $0.689\pm 0.029$ &    3 \\
R808            & $11\,160 \pm 200$  & $8.04\pm 0.05$  & $0.626\pm 0.028$ &    3 \\
BPM 24754       & $11\,070 \pm 200$  & $8.03\pm 0.05$  & $0.620\pm 0.028$ &    3 \\
G30$-$20        & $11\,070 \pm 200$  & $7.95\pm 0.05$  & $0.578\pm 0.026$ &    3 \\
PG 1149$+$058   & $10\,980 \pm 181$  & $8.10\pm 0.048$ & $0.660\pm 0.027$ &    1 \\
\hline\hline
\label{masas}
\end{tabular}}\\
{\footnotesize  
References:  $^{(1)}$Voss  et  al.   (2006), $^{(2)}$Fontaine  et  al.
(2003).  $^{(3)}$Bergeron  et al.   (2004),  $^{(4)}$Gianninas et  al.
(2006),   $^{(5)}$Gianninas  et   al.    (2005),  $^{(6)}$Homeier   et
al. (1998),$^{(7)}$Silvotti et al. (2005).

Note:  the values of  the stellar  mass marked  with $^{*}$  have been
derived  by extrapolation from  our evolutionary  model grid,  and so,
they are uncertain.}

\end{table}

In the lower panel of  Fig.  \ref{chemical_bruntv} we show the spatial
run of the logarithm  of the squared Brunt-V\"ais\"al\"a frequency for
models  with  $M_*=  0.609  M_{\odot}$  and different  values  of  the
thickness of  the H envelope for  $T_{\rm eff} \approx  12\,000$ K. In
the upper panel,  we plot the internal chemical  stratification of the
models for the  main nuclear species.  The figure  emphasizes the role
of  the chemical interfaces  on the  shape of  the Brunt-V\"ais\"al\"a
frequency.  In  fact, each  chemical transition region  produces clear
and distinctive features in  $N$, which are eventually responsible for
the mode-trapping properties of the  models. In the core region, there
are  several peaks  at  $-\log(q) \approx  0.4-0.5$  (where $q  \equiv
1-M_r/M_*$) resulting from steep  variations in the inner C-O profile.
The step shape of the C  and O abundance distribution within the core,
which is  due to  the occurrence of  extra mixing episodes  beyond the
fully convective  core during  central helium burning,  constitutes an
important   source   of  mode-trapping   in   the   core  region   ---
``core-trapped'' modes (see C\'orsico  \& Althaus 2006).  The extended
bump in $N^2$ at $-\log(q)  \approx 1-2$ is another relevant source of
mode-trapping. This  feature is caused  by the chemical  transition of
He,  C  and O  resulting  from nuclear  processing  in  prior AGB  and
thermally-pulsing AGB  stages. It  is worth noting  that the  shape of
this transition is affected by diffusion processes which are operative
at these  evolutionary stages. Finally,  there is the  He/H transition
region, which  is also another source  of mode trapping,  in this case
associated with modes trapped in the outer H envelope.

We have performed pulsation calculations on about $(11 \times 7 \times
200)  = 15\,400$  DA white  dwarf models.   In this  account,  we have
considered the number of stellar mass values ($11$), the 
number of  thicknesses of  the H  envelope for  each sequence
($\approx 7$),  and  the  number  of models  ($\approx 200$)  with  effective
temperature in the interval $14\,000-9\,000$ K, respectively. For each
model, adiabatic pulsation $g$-modes with  $\ell= 1$ and $\ell= 2$ and
periods in the  range $80-2000$ s have been  computed.  This range of
periods corresponds  (on average)  to $1 \lesssim  k \lesssim  50$ for
$\ell= 1$ and $1 \lesssim k \lesssim 90$ for $\ell= 2$.  So, more than
$\sim 2 \times 10^6$ adiabatic pulsation periods have been computed in
this work.

\begin{figure*}
\begin{center}
\includegraphics[clip,width=450pt]{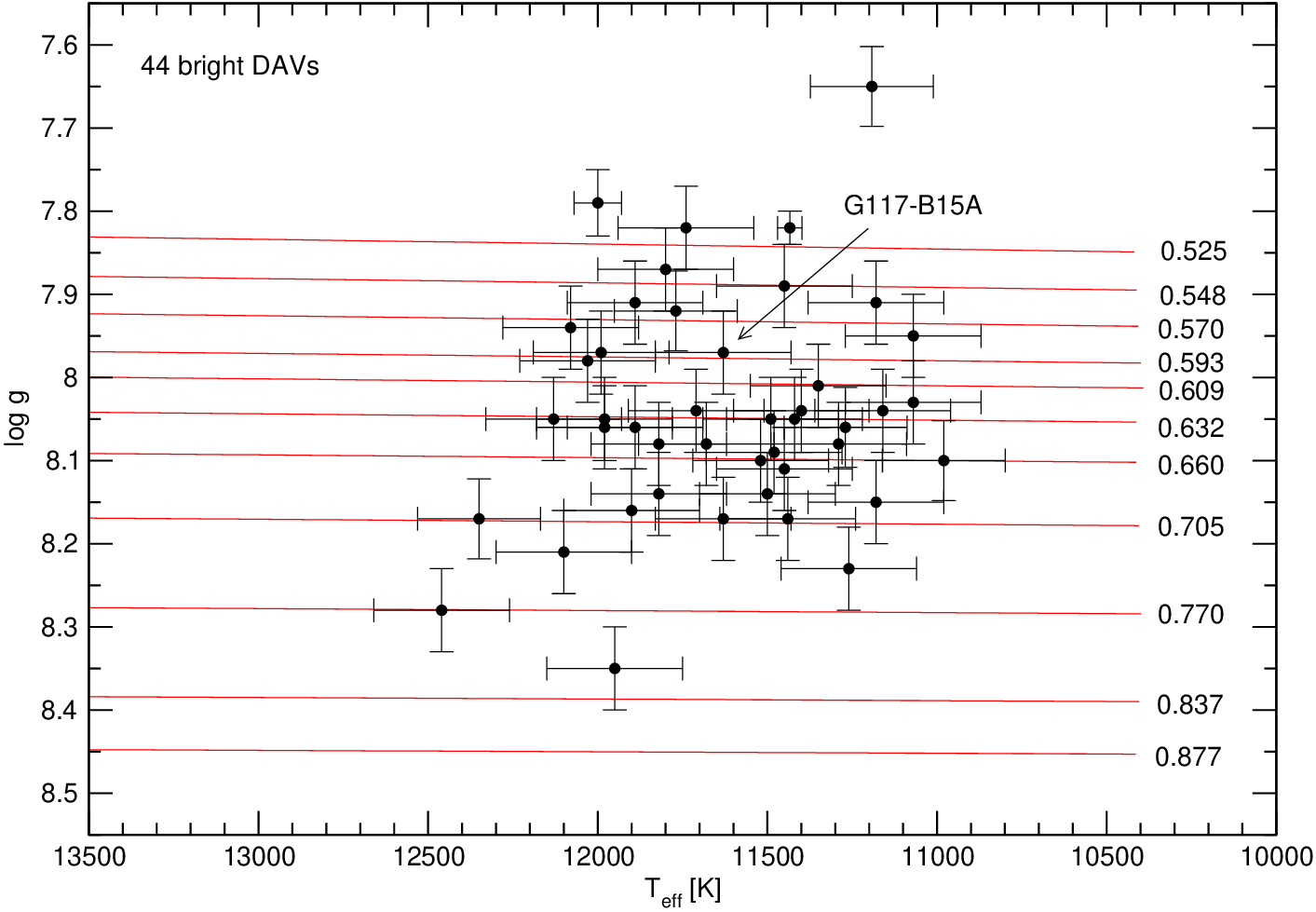}
\caption{The location of  the 44 ZZ Ceti stars  analysed in this paper
  in the $\log g- T_{\rm eff}$ plane.  The lines correspond to our set
  of  DA white  dwarf  evolutionary tracks  with  thick (canonical)  H
  envelope thickness.}
\label{gteff}
\end{center}
\end{figure*}

\section{Asteroseismological fits}
\label{astero-method}

We  search for  an  asteroseismological model  that  best matches  the
pulsation periods of our target stars.  To this end, we seek the model
that minimizes a quality function defined simply as the average of the
absolute differences  between theoretical and  observed periods (e.g.,
Bradley 1998):

\begin{equation}
\Phi = \Phi(M_*, M_{\rm H},
T_{\rm eff})= \frac{1}{N}\sum_{i=1}^N |\Pi_k^{\rm th}-\Pi_i^{\rm obs}|,
\label{phi}
\end{equation}

\noindent  where $N$  is the  number  of the  observed periods  in
the star under study.  We also have considered the quality function 
defined as (e.g., C\'orsico et al. 2009):

\begin{equation}
\chi^2= \chi^2(M_*, M_{\rm H}, T_{\rm eff})= \frac{1}{N}\sum_{i=1}^N 
\min[\Pi_k^{\rm th}-\Pi_i^{\rm obs}]^2.
\label{chi}
\end{equation}

\noindent Finally, we employ the following merit function 
(e.g., Castanheira \& Kepler 2008):

\begin{equation}
\Xi= \Xi(M_*, M_{\rm H}, T_{\rm eff})= \sum_{i=1}^N 
\sqrt{ \frac{[\Pi_k^{\rm th}-\Pi_i^{\rm obs}]^2\
 A_i}{\sum_{i=1}^N A_i}}, 
\label{s}
\end{equation}

\noindent  where the  amplitudes $A_i$  are  used as  weights of  each
observed period.   In this way, the  period fit is  more influenced by
modes with large amplitudes than by the ones with low amplitudes.

 In  the  asteroseismological  analysis  of this  work,  we  have
employed the  three quality  functions $\Phi$, $\chi^2$,  and $\Xi$,
defined  by  Equations   (\ref{phi}),  (\ref{chi}),  and  (\ref{s}),
respectively.   Since generally  these functions  lead to  very
similar results, we shall describe the quality of our period fits in
terms  of the  function $\Phi  = \Phi(M_*,  M_{\rm  H},T_{\rm eff})$
only.  The effective temperature, the  stellar mass and the mass of
the H envelope of our DA white dwarf models are allowed to vary in the
ranges:  $14\,000  \gtrsim  T_{\rm  eff}  \gtrsim  9\,000$  K,  $0.525
\lesssim  M_* \lesssim  0.877 M_{\odot}$,  $-9.4  \lesssim \log(M_{\rm
  H}/M_*) \lesssim  -3.6$, where the  ranges of the values  of $M_{\rm
  H}$  are   dependent  on  $M_*$  (see   Table\ref{table1}  and  Fig.
\ref{grid_mh}). For simplicity, the mass  of He has been kept fixed at
the  value  predicted  by   the  evolutionary  computations  for  each
sequence. As we discussed in Sect. \ref{hecontent}, the mass of the He
content is  not expected to  be substantially smaller  (say $100-1000$
times)  than  predicted by  our  modeling.   For  this not  too  large
uncertainty in the He content,  only a weak dependence of the $g$-mode
adiabatic pulsation periods  on the value of $M_{\rm  He}$ is expected
(Bradley  1996),  at variance  with  what  happens  with $M_{\rm  H}$.
Finally, artificially changing  the He mass of our  models would imply
moving the  triple transition  C-O/He, which should  introduce serious
and undesirable  artificial changes in  the chemical structure  of the
models.  The  shape of the  C-O chemical profile  at the core  and the
central abundances of O and C  have been also kept fixed according the
predictions of  the evolution during  the central He burning  stage of
the progenitors.  Finally, the thicknesses  of the C-O/He and the He/H
chemical transition  regions have also  been kept fixed at  the values
dictated by time-dependent element diffusion.

\section{Stars analysed and results}
\label{results}

We have carried out asteroseismological fits for a set of 44 bright ZZ
Ceti stars, the atmospheric parameters of which are shown in columns 2
and  3  of Table  \ref{masas}. In this Table, the stars have been 
sorted by decreasing $T_{\rm eff}$.  The location of the studied stars
in the $\log g - T_{\rm  eff}$ plane is displayed in Fig.  \ref{gteff}
along with  our evolutionary  tracks.  We defer  to a future  work the
study of the fainter ZZ Ceti stars discovered within the SDSS. Most of
these stars have  been included in the study  of Castanheira \& Kepler
(2009).

In  this section  we present  the results  of  our asteroseismological
inferences.  Because  G117$-$B15A is the benchmark the  ZZ Ceti class,
we  will devote  the complete  section \ref{g117b15a}  to  describe in
detail the results of  our asteroseismological analysis for this star,
including a discussion of our  findings, and defer the presentation of
results for the  whole sample of the analysed  stars to the subsequent
section. Before  going to  the description of  our asteroseismological
results, we briefly examine below the spectroscopic masses derived for
the studied DAV stars and how  the average value fits to the mean mass
of DA white dwarfs reported by recent works.

\subsection{Spectroscopic masses}
\label{spectroscopic-mass}

The spectroscopic masses of the 44 ZZ Ceti stars studied in this work
are shown  in column  4 of Table  \ref{masas}. They have  been derived
simply  interpolating from  the tracks  in the  $\log g-  T_{\rm eff}$
diagram given the  values of $\log g$ and  $T_{\rm eff}$ inferred from
spectroscopic analysis.   The mean  value of the  spectroscopic masses
for our sample  of DAV stars is $\langle  M_*\rangle_{\rm spec}= 0.630
\pm 0.028  M_{\odot}$.  It is  interesting to compare this  value with
the  average mass  of DA  (pulsating and  not pulsating)  white dwarfs
according to  recent studies.  Our  value is somewhat higher  ($\sim 4
\%$) than  the value reported  by Kepler et  al.  (2010) for  DA white
dwarfs on the basis of a large sample of 1505 stars of the SDSS (DR4),
$\langle M_*\rangle_{\rm DA}= 0.604  \pm 0.003 M_{\odot}$, and in
agreement  with the  recent  determination of  Falcon  et al.  (2010),
$\langle  M_*\rangle_{\rm   DA}=  0.647^{+0.013}_{-0.014}  M_{\odot}$,
obtained from the gravitational redshift determination of 449 DA white
dwarfs, and  that of Tremblay et al. (2011), $\langle
M_*\rangle_{\rm DA}= 0.613 M_{\odot}$, using  1089 DAs from DR4 of the
SDSS.

\subsection{The archetypal ZZ Ceti star G117$-$B15A}
\label{g117b15a}

For this
star, we initially computed the merit functions through our model grid
by assuming that the harmonic  degree of the three observed periods of
G117$-$B15A is $\ell= 1$  from the outset.  Somewhat disappointing, we
did  not  find  any stellar  model  of  the  basic grid  that  matched
simultaneously the  three observed periods.  By  closely examining our
results,  we discovered  that  a good  period  fit could  be found  by
considering additional values of $M_{\rm H}$ near $10^{-6} M_*$ in the
sequence with  $M_*= 0.593  M_{\odot}$ at approximately  $T_{\rm eff}=
12\,000$  K.  Hence,  we  computed several  additional sequences  with
different  values   of  $M_{\rm  H}$  until  a   best-fit  model  with
$\log(M_{\rm  H}/M_*)=  -5.903$ was  found.   The characteristics  and
periods of the best-fit model are shown in row 1 (model $1$) of 
Table \ref{table2}. The
period at 215 s  has a radial order $k= 2$.  Note  that the fit to the
main period  is excellent  ($|\Delta|\equiv |\Pi^{\rm obs}  - \Pi^{\rm
  th}|= 0.015$ s), although the  fits to the remainder two periods are
not  as good. The  global fit,  characterized by  $\Phi= 1.729$  s, is
still  very satisfactory.  We repeated  our computations  but assuming
$\ell= 1$ for the 215 s mode at the outset, and allowing the other two
periods to  be associated with $\ell=  1$ or $\ell= 2$.  We arrived at
the same asteroseismological solution.

\begin{figure}
\begin{center}
\includegraphics[clip,width=250pt]{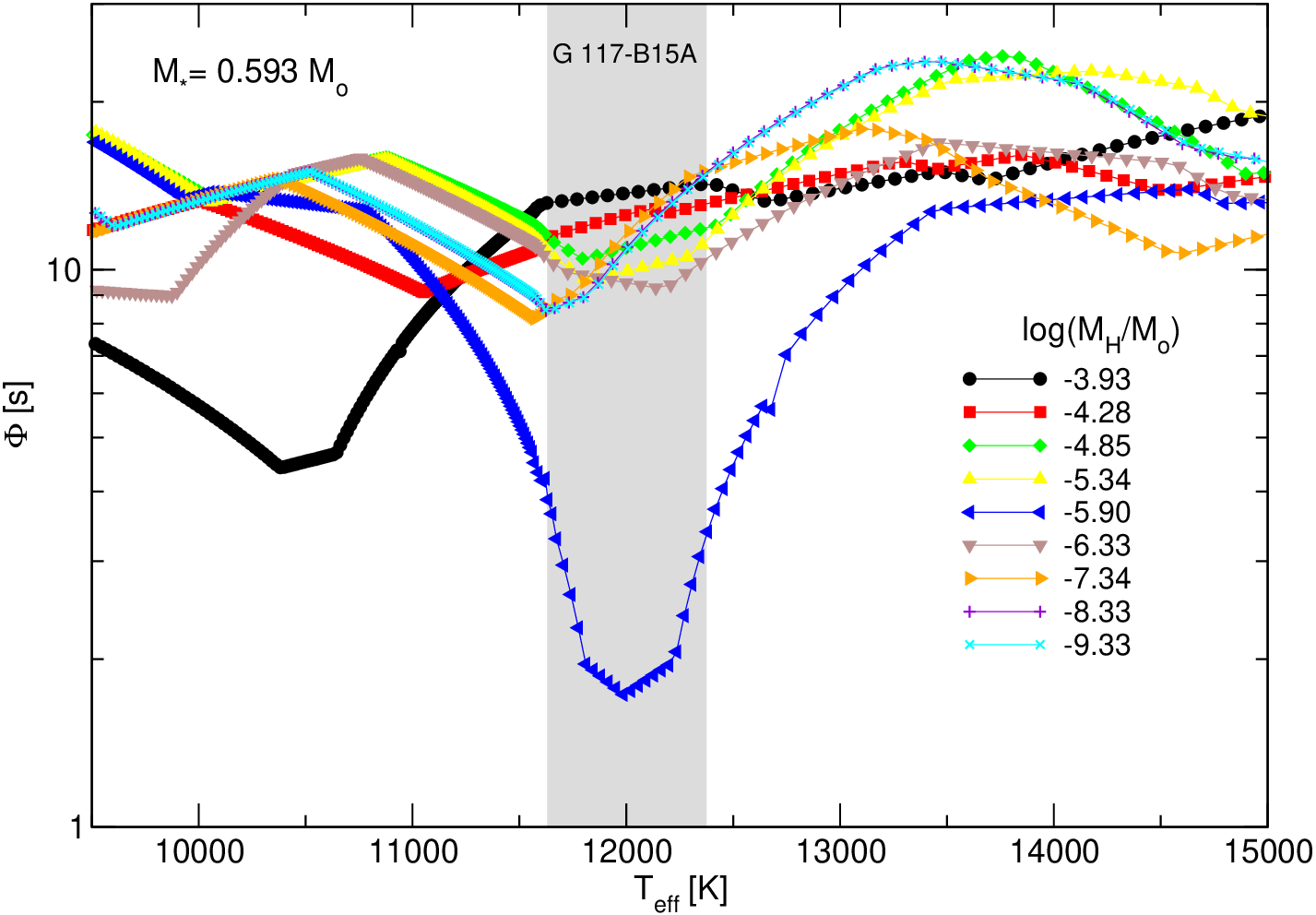}
\caption{The quality  function $\Phi(M_*, M_{\rm H},  T_{\rm eff})$ in
  terms of $T_{\rm eff}$ for the different  values of the
  hydrogen  thickness  (shown  with  different  colors  and  symbols) and
  a stellar mass of $M_*= 0.593 M_{\odot}$.  The gray
  strip  correspond  to   the  effective  temperature  of  G117$-$B15A
  according  to  spectroscopic   analysis.   Note  in  particular  the
  location of the best-fit solution  (the minimum of $\Phi$ at $T_{\rm
    eff} \sim 12\, 000$ K), corresponding to the model 1 
   ($\ell= 1$, $k= 2, 3, 4$) in Table \ref{table2}.}
\label{fi}
\end{center}
\end{figure}

In Fig.   \ref{fi} we plot  the function $\Phi(M_*, M_{\rm  H}, T_{\rm
  eff})$ in  terms of  the effective temperature  for the  different H
envelope thicknesses  corresponding to  the sequence with  $M_*= 0.593
M_{\odot}$.   Clearly  notorious  is  the existence  of  the  best-fit
solution at $T_{\rm  eff} \sim 12\, 000$ K  and $M_{\rm H}/M_* \approx
1.25  \times 10^{-6}$.   Apart from  the best-fit  solution,  there is
another minimum  at $T_{\rm eff} \sim  10\, 380$ K  and $M_{\rm H}/M_*
\approx 1.17  \times 10^{-4}$, where  $\Phi \approx 4.5$  s.  However,
this solution  must be discarded because its  effective temperature is
too  low as compared  with the  limits imposed  by spectroscopy  for 
G117$-$B15A.

\begin{table*}
\centering
\caption{Possible        asteroseismological       solutions       for
  G117$-$B15A.  Model  1 is the best-fit model corresponding to a family  
  of  solutions
  obtained by imposing that all  of the observed periods correspond to
  $\ell=1$ modes. Models 2 to 4 result from the assumption that the
  observed periods are associated to $\ell= 1$ or $\ell= 2$ modes.}
\begin{tabular}{ccccccccccc}
\hline
\hline
Model & $T_{\rm eff}$ & $M_*/M_{\odot}$ & $\log(M_{\rm He}/M_*)$ & $\log(M_{\rm H}/M_*)$ & $\Pi_i^{\rm obs}$   &  $\Pi_k^{\rm th}$ & $\ell$ & $k$ & $|\Delta|$ & $\Phi$ \\
 & $[$K$]$      &                &                       &                     &   $[$s$]$     &   $[$s$]$    &   &     &    $[$s$]$ & $[$s$]$\\  
\hline
1 & $11\,986$   &   $0.5932$      &    $-1.62$            &   $-5.90$          &   215.20   &    215.215   &   1  &  2    &  0.015 &   1.729   \\
  &             &                 &                       &                    &   270.46   &    273.437   &   1  &  3    &  2.977 &    \\
  &             &                 &                       &                    &   304.05   &    301.854   &   1  &  4    &  2.196 &    \\
\hline
\hline
2 & $12\,450$   & $0.6090$        &   $-1.61$             &   $-4.45$          & 215.20        &   214.947      & 2      & 6   & 0.253  & 0.177  \\
  &             &                 &                       &                    & 270.46        &   270.268      & 2      & 8   & 0.192  &   \\
  &             &                 &                       &                    & 304.05        &   304.136      & 2      & 9   & 0.086  &   \\
\hline
3 & $12\,219$   & $0.6598$        &   $-1.91$             &   $-8.33$          & 215.20        &   215.218      & 2      & 5   & 0.018  & 0.526  \\
  &             &                 &                       &                    & 270.46        &   270.406      & 1      & 3   & 0.054  &   \\
  &             &                 &                       &                    & 304.05        &   305.557      & 2      & 8   & 1.507  &   \\
\hline
4 & $11\,735$   & $0.5930$        &   $-1.62$             &   $-9.33$          & 215.20        &   214.422      & 2      & 4   & 0.778  & 0.735  \\
  &             &                 &                       &                    & 270.46        &   271.682      & 1      & 2   & 1.222  &   \\
  &             &                 &                       &                    & 304.05        &   303.846      & 2      & 7   & 0.204  &   \\
\hline
\end{tabular}
\label{table2}
\end{table*}

The  uniqueness of  the  solution  regarding the  thickness  of the  H
envelope   is   one  of   the   main   results   of  this   work   for
G117$-$B15A. However,  we warn  that in this  study we are  matching 3
observed quantities (the pulsation  periods of G117$-$B15A) by varying
just 3  structural quantities ($M_*,  M_{\rm H}$, and  $T_{\rm eff}$).
So,  it is not  unconceivable that  if we  were varying  an additional
parameter (for instance $X_{\rm O}$, $M_{\rm He}$, etc) of our models,
we  could  found multiple  asteroseismological  solutions  due to  the
ambiguity introduced by the new  parameter to be adjusted.  Our models
do not have an extra fit parameter.

We  also carried  out additional  period fits  in which  the  value of
$\ell$ for each of the theoretical periods is not fixed but instead is
obtained  as an  output  of  our period  fit  procedure, although  the
allowed  values are just  $\ell= 1$  and $\ell=  2$.  The  results are
displayed in rows 2 to 4 of Table \ref{table2}.  For these models, the
period  fits are  excellent.  In  particular, the  periods of  model 2
match the observed  periods with an average difference  of $\sim 0.18$
s. One  of the  reasons is that  are more  $\ell= 2$ modes  per period
interval.   However, for  the three  models, the  main  periodicity of
G117$-$B15A at  215 s is associated to  a $\ell= 2$ mode.   This is in
strong contradiction with the results of Robinson et al.  (1995), who
identify the 215 s period with a $\ell= 1$ mode by
means  of  time-resolved  ultraviolet  spectroscopy.  This  result  is
consistent with the further analysis of Kotak et al.  (2004). Thus, as
tempting as these solutions seem,  they must all be discarded from our
analysis.

\begin{figure*}
\begin{center}
\includegraphics[clip,width=500pt]{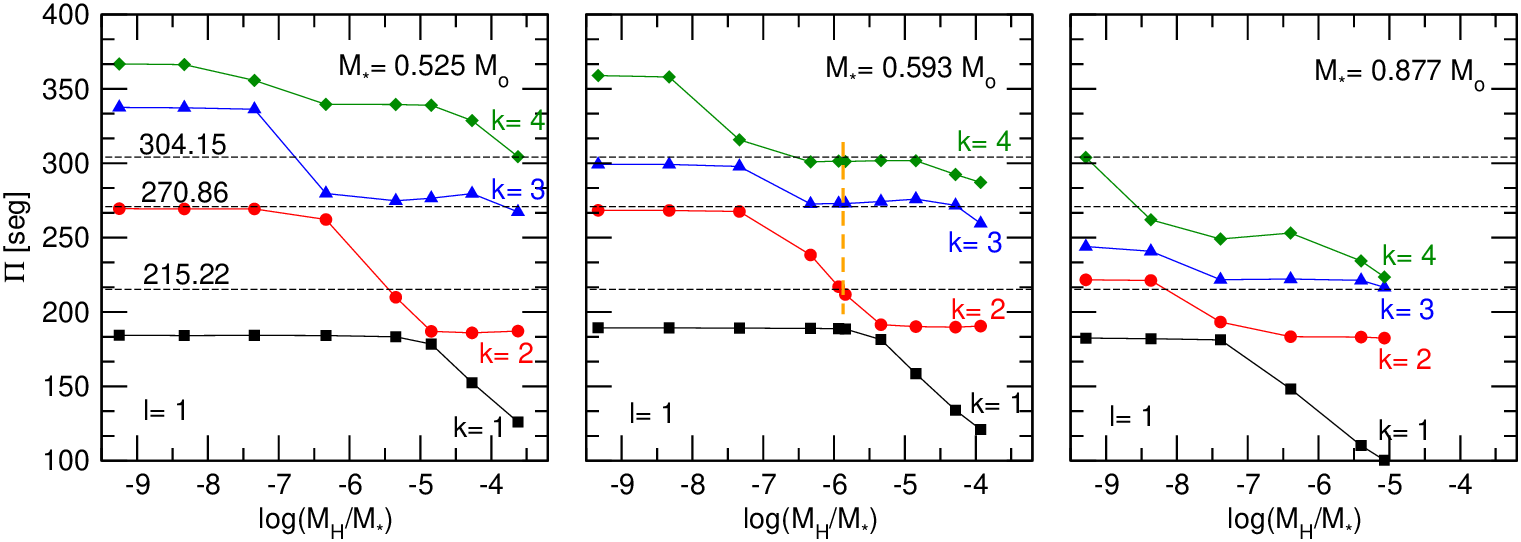}
\caption{The periods of  the modes with $\ell= 1$ and $k= 1, 2, 3$  
  and $4$ in terms
  of  the thickness  of the  H envelope  corresponding to  $M_*= 0.525
  M_{\odot}$ (left panel), $M_*=  0.593 M_{\odot}$ (middle panel), and
  $M_*=  0.877  M_{\odot}$  (right  panel). The  observed  periods  of
  G117$-$B15A are  shown with thin  horizontal dashed 
  lines. The  dashed (orange)
  line in the middle panel indicates the match between the theoretical
  periods of the asteroseismological model and the periods observed in
  G117$-$B15A.}
\label{todos}
\end{center}
\end{figure*}

\subsubsection{Estimation of the internal uncertainties}

We   have   assessed   the   uncertainties   in   the   stellar   mass
($\sigma_{M_*}$),  the thickness  of the  H  envelope ($\sigma_{M_{\rm
    H}}$)  and the effective  temperature ($\sigma_{T_{\rm  eff}}$) of
the  best-fit model by  employing the  expression (Zhang et al. 1986; 
Castanheira \& Kepler 2008):

\begin{equation}
\sigma_i^2= \frac{d_i^2}{(S-S_{\rm 0})},
\end{equation}

\noindent where $S_{\rm 0}\equiv  \Phi(M_*^{\rm 0}, M_{\rm H}^{\rm 0},
T_{\rm eff}^{\rm  0})$ is  the minimum of  $\Phi$ which is  reached at
$(M_*^{\rm 0}, M_{\rm H}^{\rm  0},T_{\rm eff}^{\rm 0} )$ corresponding
to the best-fit  model, and $S$ is the value of  $\Phi$ when we change
the parameter $i$  (in this case, $M_*, M_{\rm  H}$, or $T_{\rm eff}$)
by an amount $d_i$, keeping  fixed the other parameters.  The quantity
$d_i$  can  be evaluated  as  the  minimum step  in  the  grid of  the
parameter $i$.   We obtain the following uncertainties,  which are the
{\sl    internal}    errors    of   our    asteroseismic    procedure:
$\sigma_{M_*}\sim  0.007  M_{\odot}$,  $\sigma_{M_{\rm  H}}  \sim  0.7
\times  10^{-6}  M_*$,  and  $\sigma_{T_{\rm eff}}\sim  200$  K.   The
uncertainties in the other quantities ($L_*, R_*, g$, etc) are derived
from the uncertainties in $M_*$ and $T_{\rm eff}$.

In  Table \ref{table3},  we compare  the main  characteristics  of our
best-fit  model  with  the  observed properties  of  G117$-$B15A.   In
particular,  we include  the surface  parameters of  G117$-$B15A taken
from several spectroscopic studies.  We include also the spectroscopic
mass computed by interpolating from our evolutionary tracks.  Note the
 agreement  between the effective temperature  and gravity of
our  asteroseismological model and  the values  derived by  Koester \&
Allard (2000)  and Koester \& Holberg  (2001).  The total  mass of our
model, however, is  $3-11$ \% higher than the  values derived in those
studies.   Our model  is $\sim  350$ K  hotter than  the spectroscopic
temperature of Bergeron et al. (1995a, 2004), and about $400$ K cooler
than the  value derived  by Robinson et  al.  (1995), but  the surface
gravity and mass are in  excellent agreement with the values quoted in
both studies.

\subsubsection{Asteroseismological distance}

Since we  have the luminosity of  the best-fit model,  we can estimate
the asteroseismological distance and  parallax of G117$-$B15A by means
of  the relation  $\log d  [{\rm pc}]=  \frac{1}{5} \left(m_{\rm  V} -
M_{\rm  V} +  5  \right)$,  where $M_{\rm  V}=  M_{\rm bol}-BC$.   The
bolometric magnitude, $M_{\rm bol}$,  can be computed as $M_{\rm bol}=
M_{\rm  bol}(\odot) - 2.5  \log(L_*/L_{\odot})$, being  the bolometric
magnitude  of the  Sun $M_{\rm  bol}(\odot)= +4.75$  (Allen  1973). By
using  $BC= -0.611$  (Bergeron et  al. 1995a)  and $m_{\rm  V}= 15.50$
(Bergeron et al. 1995b), we obtain a distance $d= 60.3\pm 2.5$ pc, and
a parallax  $\pi= 16.6 \pm 0.8$  mas, in excellent  agreement with the
inference of Bradley (1998) ($\pi= 16.5$ mas).  The distance estimated
from optical, IUE,  and HST spectra is  $58 \pm 2$ pc, $59  \pm 5$ pc,
and $67 \pm 9$ pc,  respectively. Holberg et al. (2008) derive
a distance of $57.68 \pm 0.60$ pc. Our seismological parallax is larger
than the trigonometric value  extracted from the Yale Parallax Catalog
(van  Altena et  al.  1994)  of $10.5\pm  4.2$ mas.   In order  for the
asteroseismological parallax to be compatible with the trigonometric one,
the mass of the asteroseismological model should be as low as $\approx
0.35  M_{\odot}$!  We  can  safely  discard a  low  stellar  mass  for
G117$-$B15A   from   spectroscopy.   Then,   we  conclude   that   the
trigonometric parallax  must be more  uncertain than quoted,  and that
the asteroseismological parallax is robust.

\begin{table*}
\centering
\caption{Characteristics  of  G117$-$B15A  and  of  our  seismological
  model. The  quoted uncertainties in the seismological  model are the
  internal errors of our  period-fit procedure. The progenitor star of
  the asteroseismological model star has  a stellar mass of $M_*= 1.75
  M_{\odot}$ at the ZAMS.}
\begin{tabular}{lccccc}
\hline
\hline
 Quantity             & Robinson et al. &   Koester \& Allard & Koester  \& Holberg & Bergeron et al.   &   Our seismological  \\
                      & (1995)          &   (2000)            &   (2001)            &     (1995a, 2004)  &               model  \\          
\hline
$T_{\rm eff}$ [K]      &$12\,375\pm 125$ &   $11\,900\pm 140$  &  $12\,010\pm 180$    & $11\,630\pm 200$ &   $11\,985 \pm 200$  \\
$M_*/M_{\odot}$        &$0.591 \pm 0.031$&   $0.534 \pm 0.072$ &  $0.575 \pm 0.092$     & $0.589 \pm 0.026$  &   $0.593 \pm 0.007$   \\
$\log g$              & $7.97\pm 0.06 $ &   $7.86\pm 0.14 $   &  $7.94\pm 0.17 $     & $7.97\pm 0.05 $  &   $8.00\pm 0.09$   \\
$\log(R_*/R_{\odot})$  &    ---          &       ---           &      ---             &  ---             &   $-1.882\pm 0.029$  \\   
$\log(L_*/L_{\odot})$  &    ---          &       ---           &      ---             &  ---            &    $-2.497 \pm 0.030$  \\
$M_{\rm He}/M_*$       &    ---          &       ---           &      ---             &  ---            &   $2.39  \times 10^{-2}$  \\
$M_{\rm H}/M_*$        &    ---          &       ---           &      ---             &  ---            &   $(1.25\pm 0.7) \times 10^{-6}$   \\
$X_{\rm C},X_{\rm O}$ (center)   &    ---          &       ---           &      ---             &  ---            &   $0.28, 0.70$   \\
\hline
\end{tabular}\\
{\footnotesize Note 1: the values  of the spectroscopic mass quoted in
  columns 2 to  5 have been computed by interpolating  from our set of
  evolutionary tracks (see Fig.  \ref {gteff}) using the corresponding
  values of $\log g$ and $T_{\rm eff}$. 

  Note 2: Robinson et al. (1995)
  use MLT$/\alpha=  1$ model atmospheres, while the  other use 
MLT$/\alpha=  0.6$, hence they obtain lower $T_{\rm eff}$s.}
\label{table3}
\end{table*}

\subsubsection{Discussion}

All the  previous asteroseismological studies  on G117$-$B15A (Bradley
1998, Benvenuto et al.  2002, Castanheira \& Kepler 2008, Bischoff-Kim
et  al.  2008a)  report an  ambiguity of  the solutions  regarding the
thickness of the  H envelope of this star. In  those studies, a family
of thin envelope solutions is  obtained for a identification $k= 1, 2,
3$, whereas a second family of solutions of thick envelopes is derived
if  $k=  2, 3,  4$.   In  contrast,  our asteroseismological  analysis
strongly points to a single  solution regarding the thickness of the H
envelope,  with  a value  of  $\log(M_{\rm  H}/M_*)  \sim -5.9$,  that
corresponds to the  identification $k= 2, 3, 4$,  which was associated
to  thick H  envelopes in  the  previous studies.   The degeneracy  of
solutions is solved for the first time by our results.  The reason for
the uniqueness of the solution in our computations is that, regardless
of the value  of the stellar mass, temperature, or  thickness of the H
envelope, it is impossible to find a model whose mode with $k = 1$ has
a  period close  to 215  s.   This preclude  us to  find any  possible
asteroseismological model with the  identification $k= 1, 2, 3$.  This
is shown  in Fig. \ref{todos}, where  we plot the periods  in terms of
the thickness of the H envelope for models with $T_{\rm eff} \sim 12\,
000$ K and $M_*= 0.525 M_{\odot}$ (left panel), $M_*= 0.593 M_{\odot}$
(middle  panel), and  $M_*= 0.877  M_{\odot}$ (right  panel).   In the
middle panel,  the best-fit model  is indicated with a  dashed (brown)
line.  Clearly, the  period for the mode with $k= 1$  in our models is
always  very short  in comparison  with the  shortest period  shown by
G117$-$B15A.   We mention  that we  have also  computed  an additional
evolutionary sequence  with the  same characteristics as  our best-fit
model, but  with a  thinner H envelope  than that considered  in Table
\ref{table1} ($\log (M_{\rm H}/M_*) <  -9.33$). Even in this case, the
pulsation periods  exhibit the  same trend shown  in the  middle panel
Fig.   \ref{todos},  with the  period  of  the  $k= 1$  mode  markedly
departed  from 215  s, and  as a  result, we  are not  able to  find a
thin-envelope solution.

It should  be kept in mind,  however, that we could  run into multiple
solutions for G117$-$B15A  if we were to vary  an additional parameter
of our  models. For  instance, it  could be possible  that if  we were
freely  changing the  He content  ($M_{\rm  He}$) of  our models,  for
instance by adopting  a He layer mass two  order of magnitude thinner,
then the $k= 1$  period could became close to 215 s,  and so, we could
recover the two families of thin and thick He envelope solutions found
in    the    previous    studies.     However,   as    discussed    in
Sect. \ref{hecontent},  such low $M_{\rm He}$ values  are difficult to
conceive from stellar evolution calculations.

A distinctive feature shown in  Fig.  \ref{todos} is the presence of a
behavior reminiscent to the  well known ``avoided crossing'' (see also
Fig.   3  of  Castanheira \&  Kepler  2008).   When  a pair  of  modes
experiences  avoided  crossing,  the  modes exchange  their  intrinsic
properties  (see  Aizenman  et  al.   1977). In  our  models,  avoided
crossing is produced when we vary  the thickness of the H envelope. As
a result, for certain values  of $M_{\rm H}$, the period spacing turns
out be very short. This effect  is more notorious for low radial order
modes. For instance, for the sequence with $M_*= 0.525 M_{\odot}$, the
period spacing  between the modes  with $k= 1$  and $k= 2$ is  of only
$\approx 8$  s if $\log(M_{\rm H}/M_*) \sim  -4.8$!  Something similar
is seen for the sequences  with $M_*= 0.593 M_{\odot}$ and $M_*= 0.877
M_{\odot}$, with  H envelopes of  $\log(M_{\rm H}/M_*) \sim  -5.3$ and
$\log(M_{\rm  H}/M_*) \sim  -7.4$,  respectively.  We  note also  that
avoided crossing is present in our  models, but to a less extent, when
we vary the effective temperature.

\begin{figure}
\begin{center}
\includegraphics[clip,width=250pt]{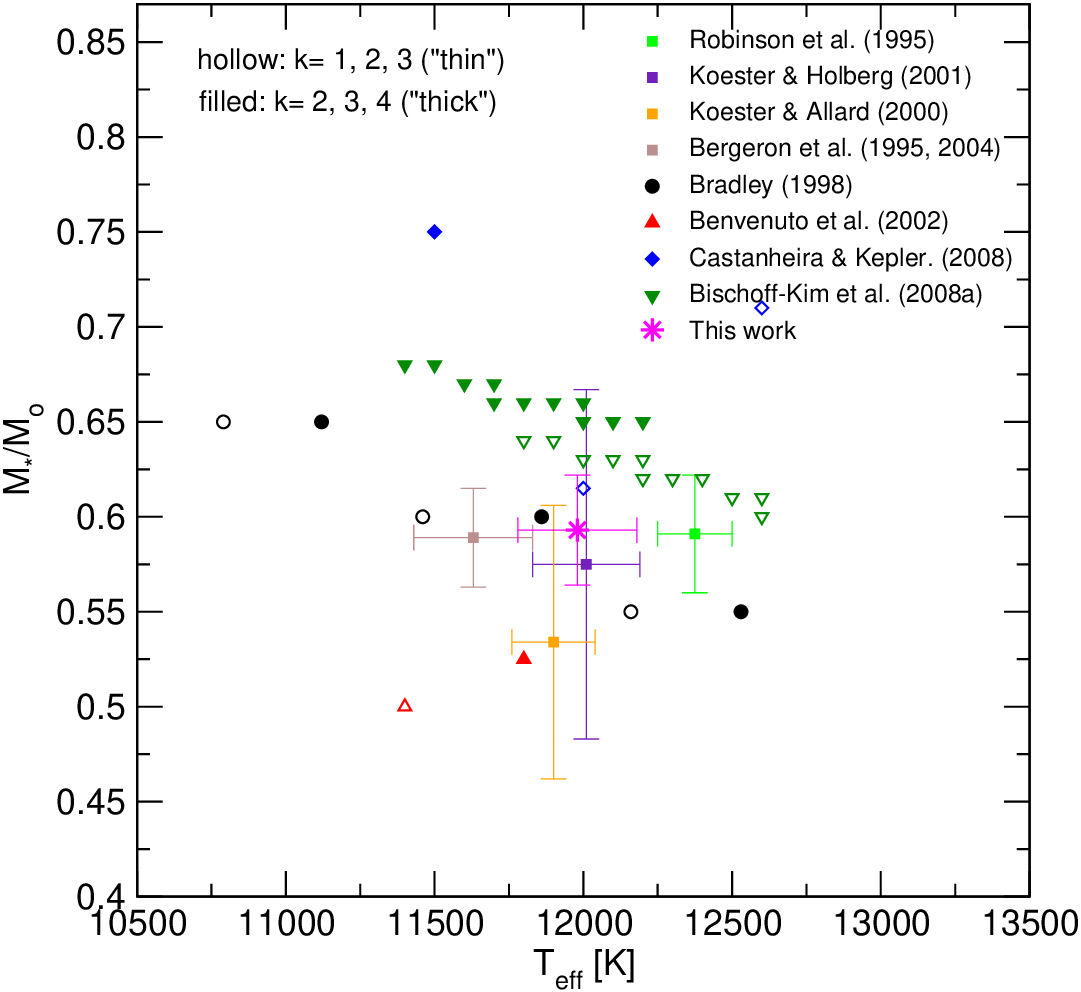}
\caption{The   location   of   the  asteroseismological   models   for
  G117$-$B15A in the plane  $T_{\rm eff}-M_*$ according to the studies
  carried out up to date and according to this work, as indicated with
  different symbols.  Empty symbols  correspond to solutions for which
  the radial order identification of the observed periods is $k= 1, 2,
  3$  (thin  H  envelopes),  and  filled  symbols  are  associated  to
  solutions for which the observed periods  have $k= 2, 3, 4$ (thick H
  envelopes).  Also included is  the location of G117$-$B15A according
  to  several  spectroscopic  studies  (filled  squares).   Note  that
  Robinson et al. (1995)  use MLT$/\alpha= 1$ model atmospheres, while
  the  other use MLT$/\alpha=  0.6$, hence  they obtain  lower $T_{\rm
    eff}$ values.}
\label{m_teff}
\end{center}
\end{figure}

\begin{figure}
\begin{center}
\includegraphics[clip,width=250pt]{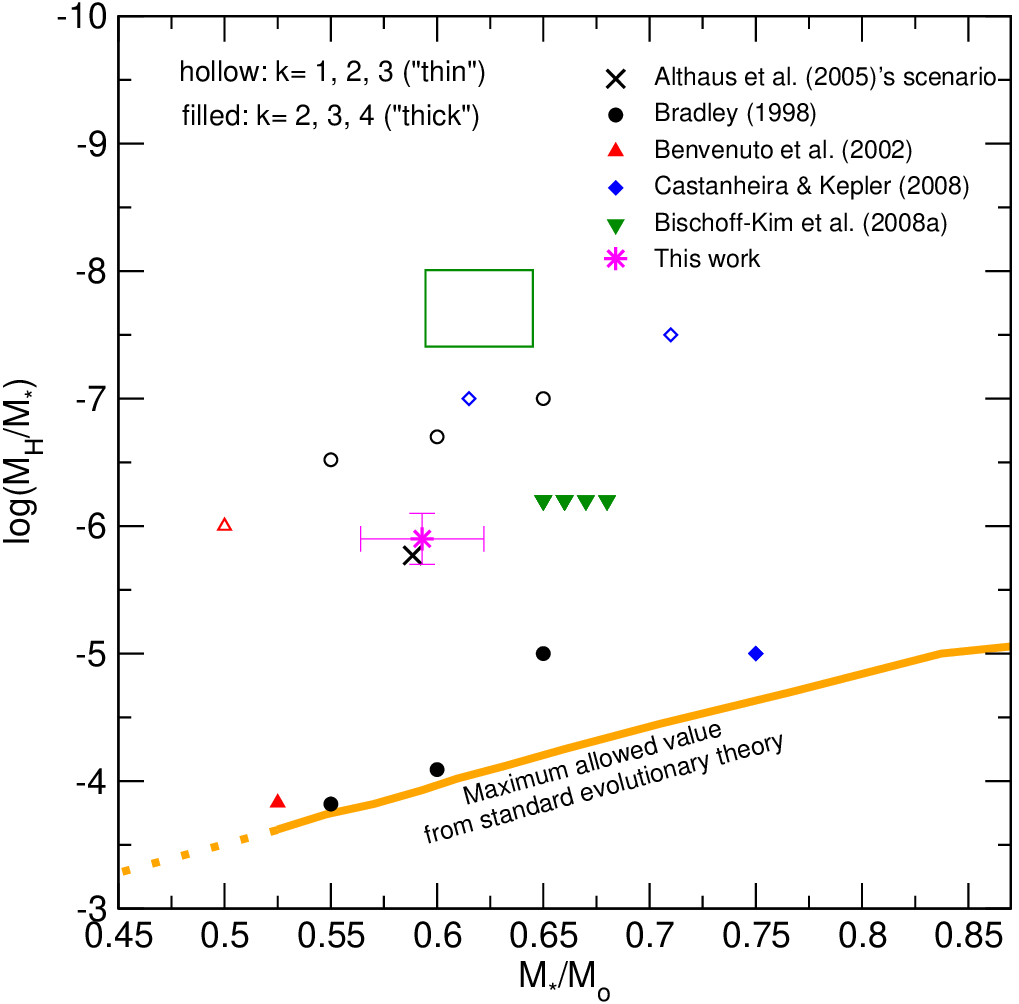}
\caption{Similar   to   Fig.    \ref{m_teff},   but   in   the   plane
  $M_*-\log(M_{\rm H}/M_*)$. The hollow green rectangle corresponds to
  the thin  solutions of Bischoff-Kim et al.  (2008a) corresponding to
  an identification  $k= 1,2,3$. In  the interests of  comparison, the
  location  of the  post-Late  Thermal  Pulse remnant  with  a thin  H
  envelope of the scenario of Althaus et al. (2005b) is also shown.}
\label{mh_m}
\end{center}
\end{figure}

In Figs.  \ref{m_teff}  and \ref{mh_m}, we display the  thin and thick
envelope solutions  with empty  and filled symbols,  respectively, for
the  asteroseismological  solutions  found  in  previous  works.   The
location of  our asteroseismological model is depicted  with a (magenta)
star  symbol.  Note  that Bischoff-Kim  et al.  (2008a)  found several
equally  valid  asteroseismological  models  with  thin  and  thick  H
envelopes   for  G117$-$B15A.   Fig.    \ref{m_teff}  shows   a  clear
correlation  between the  mass and  the effective  temperature  of the
solutions  in   the  studies   of  Bradley  (1998),   Bischoff-Kim  et
al. (2008a)  and Castanheira \&  Kepler (2008): cooler  solutions have
larger masses.  The  opposite trend is exhibited by  the two solutions
of Benvenuto et al. (2002).   All the solutions of Bischoff-Kim et al.
(2008a)  and  two  solutions  of  Castanheira  \&  Kepler  (2008)  are
substantially more massive than the best-fit models of Bradley (1998),
Benvenuto et  al. (2002), and  our own asteroseismological  model, and
also  than the  estimations of  the stellar  mass of  G117$-$B15A from
spectroscopic studies.

The  existence of  two separate  families of  solutions  regarding the
thickness  of the  H envelope,  as predicted  by previous  studies, is
clearly  emphasized  in  Fig.   \ref{mh_m}.   Here, it  is  notable  a
correlation between $M_{\rm H}$ and $M_*$, according to which the more
massive asteroseismological models have thinner envelopes.  Curiously,
this  trend  is  in  line   with  the  predictions  of  the  canonical
evolutionary computations (Althaus et al. 2010b), which are shown with
a thick (orange) line that  connects the maximum values of $M_{\rm H}$
for  different  stellar  masses.   The  figure also  shows  a  notable
agreement  between the  thin and  thick solutions  of  Bradley (1998),
Benvenuto  et  al.  (2002),  and  Castanheira  \&  Kepler (2008).   In
contrast, the solutions of Bischoff-Kim et al.  (2008a) (both thin and
thick)  appear  shifted  toward  smaller  values  of  the  H  envelope
thickness.  In  this context, our single  seismological solution seems
to  be more  nearly compatible  with the  family of  thin  H envelopes
(although  with  a   mode  identification  typical  of  thick-envelope
solutions) than with the group  of thick envelopes.

Although our new value for  the H envelope thickness of G117$-$B15A is
significantly  lower than  the  canonical value  predicted by  stellar
evolution (roughly  2 orders of  magnitude thinner), it is  in perfect
agreement with what could be expected from the LTP scenario.  
In this scenario, also called AGB final thermal pulse or 
AFTP scenario, a final helium shell flash is experienced by a star 
shortly after the departure
from the AGB (Bl\"ocker 2001). During a LTP,  not all the hydrogen  
is burnt, in contrast to 
post-AGB stars that experience a very late thermal pulse (a born-again episode), 
but part is diluted by surface convection and mixed inwards with the underlying
intershell region formerly enriched in helium, carbon and oxygen.
Althaus  et  al.   (2005b)  have  explored  the  possibility  that  an
initially $2.7  M_{\odot}$ star experiencing  a LTP shortly  after the
departure from the thermally pulsing AGB could reach the final cooling
branch  with a  H  envelope substantially  smaller  than predicted  by
standard  stellar evolution.   They found  that most  of  the original
H-rich material  of the post-AGB  remnant is burnt after  the post-LTP
evolution, when the star returns  to the high $T_{\rm eff}$ regime for
the second  time, resulting in a  white dwarf remnant  of $M_*= 0.5885
M_{\odot}$ with a value of the H envelope thickness of $M_{\rm H}= 1.7
\times  10^{-6}  M_*$.  Very  interestingly,  our  best-fit model  for
G117$-$B15A and the  DA white dwarf model resulting  from the scenario
proposed by  Althaus et  al. (2005b) are  located roughly at  the {\sl
  same}   place   in   the   plane  $M_*-\log(M_{\rm   H}/M_*)$   (see
Fig.  \ref{mh_m})\footnote{We  warn, however,  that  both models  have
  different  internal chemical  structure,  in particular  due to  the
  presence  of extra  chemical  structure and  appreciable amounts  of
  $^{14}$N at the base of the He buffer in the post-LTP DA white dwarf
  model of Althaus et al.  (2005b)  (see the panel D of Figure 2 of 
  Miller Bertolami et al. 2005).}.
Therefore, our study reinforces the validity of the results of Althaus
et  al.   (2005b)  about the  existence  of  DA  white dwarfs  with  H
envelopes substantially  thinner than the canonical  value, and suggests
that G117$-$B15A could be the  descendant of a  progenitor star that
experienced a LTP episode before reaching the final cooling branch.

\subsection{The set of 44 ZZ Ceti stars}
\label{zzcetis}

Here, we present  the asteroseismological analysis for the  44 ZZ Ceti
stars  listed  in Table  \ref{masas},  G117$-$B15A  included.  In  the
second and third  columns of Table 5 we show  the observed periods and
amplitudes, respectively. These values are extracted from the works of
Castanheira \&  Kepler (2008, 2009), unless  indicated otherwise.  The
fourth column of Table 5  shows the theoretical periods of the adopted
asteroseismological model  for each star, whereas the  fifth and sixth
columns include  the $\ell$- and  $k$-identification, respectively, of
each pulsation mode.  The seventh column shows the absolute difference
between  observed  and  theoretical  periods, and  the  eighth  column
indicates   the   value   of   the   quality   function   defined   by
Eq. (\ref{phi}).

Below,  we  describe  the  general  criteria  adopted  to  choose  the
asteroseismological model for each star.

\subsubsection{Criteria used in the fits}

Usually, when  performing period-to-period fits  to ZZ Ceti  stars, we
found multiple  seismological solutions, that is,  many stellar models
that nearly reproduce  the periods observed in a  given DAV star.  So,
in  order to  isolate  a single  asteroseismological  model among  the
several  possible  and equally  valid  ones,  we  must apply  some
criteria:

\begin{itemize}

\item First, we  looked for the models associated  to the lowest value
  of the  quality functions, thus  ensuring that the  observed periods
  are closely matched by the theoretical ones.

\item When  possible, we  used the external  $\ell$-identifications of
  the observed periods according to studies that employ the high-speed
  photometry   method  (see,   e.g.,  Robinson   et  al.   1995),  the
  time-resolved ultraviolet spectroscopy  method (see, e.g., Kepler et
  al. 2000)  or the time-resolved optical  spectroscopy approach (see,
  e.g., Clemens et al. 2000).

\item When  several families of  solutions were found, we  elected the
  models  with  values of  $T_{\rm  eff}$ and  $\log  g$  as close  as
  possible to the  spectroscopic ones. In this way,  we guarantee that
  the surface parameters of  the asteroseismological solutions are not
  in conflict with observations.

\item Among possible asteroseismological solutions with similar values
  of the quality  function, we prioritized the solutions  that fit the
  largest  amplitude  modes   with  theoretical  modes  having  $\ell=
  1$. This  is because  the well-known property  that $\ell=  1$ modes
  exhibit substantially larger amplitudes than $\ell= 2$ ones, because
  geometric cancellation effects  become increasingly severe as $\ell$
  increases (Dziembowski 1977).

\item In  the cases in which  several modes had  similar amplitudes in
  the power spectrum,  we gave more weight to  stellar models that fit
  those  periods  with  theoretical  periods having  the  same  $\ell$
  value.  In  this way,  we  are  assuming  that two  eigenmodes  with
  different  values of  the  harmonic degree  $\ell$  usually 
  should not have  similar amplitudes.

\item For a given star showing a large number of modes, we favored the
  seismological solutions  that fit to observed periods  with a larger
  number of $\ell= 1$ than $\ell=  2$ modes.  This is because there is
  more chance to observe $\ell= 1$ modes than $\ell= 2$ modes.

\item In the opposite case, for stars exhibiting just a single period,
  we employed only the set of  $\ell= 1$ periods to perform the period
  fit.  Then, we chosen the asteroseismological model by searching for
  that model having the minimum  value of the quality function, and we
  restricted  the  solutions by  using  the spectroscopic  constraints
  ($T_{\rm eff}$ and $\log g$), if necessary.

\end{itemize}

\begin{table*}
\centering
\caption{Periods observed in the sample of 44 bright ZZ Ceti stars 
studied in this work and the corresponding theoretical periods 
and ($\ell, k$)-identification 
of our asteroseismological models.}
\begin{tabular}{lccccccc}
\hline 
\hline
Star  &  $\Pi^{\rm obs}$ & $A$          &  $\Pi^{\rm th}$  &  $\ell$ &  $k$ & $|\Delta|$  & $\Phi$ \\
      &  $[$s$]$        & $[$mma$]$  &  $[$s$]$     &         &      &  $[$s$]$  &   $[$s$]$ \\ 
\hline
\hline
HS 1531$+$7436 & 112.50   & $\cdots$ &  112.499   &  1 &    1   & 0.001  & 0.001 \\
\hline 
GD 244 $^{(1)}$  &    202.98  &    4.04  &  195.973   &  2 &  5 & 7.007  & 2.165 \\          
       	        &    256.56  &   12.31  &  257.215   &  1 &    3   & 0.665  & \\
	        &    294.60  &    4.85  &  296.820   &  2 &    9   & 2.220  &  \\
                &    307.13  &   20.18  &  306.283   &  1 &    5   & 0.847  & \\
                &    906.08  &    1.72  &  906.176   &  1 &   19   & 0.086  & \\
\hline 

G226$-$29       &    109.28 & $\cdots$ &  109.246  &  1 &    1    & 0.032  & 0.032 \\
\hline 
HS 0507$+$0434B &    355.80   &   24.0   &  356.737   &  1 &    6   & 0.937  & 0.778 \\    
                &    446.20   &   13.9   &  446.429   &  1 &    8   & 0.229  & \\
                &    555.30   &   16.6   &  556.767   &  1 &   11   & 1.468  & \\
                &    743.40   &    7.6   &  742.920   &  1 &   16   & 0.679  & \\
\hline 
LP 133$-$144    &    209.20   &   10     &  211.247   &  1 &    2   & 2.047  & 1.256 \\
                &    305.70   &   5.3    &  304.394   &  2 &    8   & 1.306  & \\
                &    327.30   &   4.0    &  327.716   &  2 &    9   & 0.416  & \\
\hline 
EC 11507$-$1519 &    191.70   &   3.59   &  191.964   &  1 &    2   & 0.264  & 0.231\\
                &    249.60   &   7.70   &  249.798   &  1 &    4   & 0.198  & \\
\hline 
L19$-$2         &    113.80   &   2.4    &  113.313   &  2 &    2   & 0.487  & 1.224\\
                &    118.70   &   1.2    &  114.495   &  1 &    1   & 4.205  & \\
                &    143.60   &   0.6    &  143.272   &  2 &    3   & 0.128  & \\
                &    192.60   &   6.5    &  192.561   &  1 &    2   & 0.039  & \\
                &    350.10   &   1.1    &  351.359   &  1 &    6   & 1.259  & \\
\hline 
GD 66 $^{(2)}$   &    197.65  &   4.21   &  198.104   &  2 &    4   & 0.450  & 0.871\\
                &    255.87  &   3.43   &  256.137   &  2 &    6   & 0.270  & \\
                &    271.71  &  16.70   &  271.804   &  1 &    3   & 0.089  & \\ 
                &    302.77  &  11.29   &  300.102   &  1 &    4   & 2.663  & \\
\hline 
G132$-$12       &    212.70  &   4.3    &  212.703   &  1 &    2   & 0.003  & 0.003  \\
\hline 
G207$-$9        &    259.10   &  17.3    &  258.853   &  1 &    4   & 0.247  & 0.767 \\  
                &    292.00   &  49.0    &  290.379   &  2 &   10   & 1.621  & \\
                &    318.00   &  64.0    &  318.257   &  1 &    5   & 0.257  & \\
                &    557.40   &  63.4    &  556.204   &  1 &   12   & 1.376  & \\
                &    740.40   &  46.4    &  741.034   &  1 &   17   & 0.334  & \\
\hline 
G117$-$B15A     &    215.20  & 17.36    &  215.215   &  1 &    2   & 0.015  & 1.729 \\    
                &    270.46  &  6.14    &  273.437   &  1 &    3   & 2.977  & \\   
                &    304.05  &  7.48    &  301.854   &  1 &    4   & 2.196  & \\     
\hline 
MCT 2148$-$2911 &    260.80   & 12.6     &  260.798   &  1 &    4   & 0.002  & 0.002 \\
\hline
\label{ajustes}
\end{tabular}
\end{table*}

\begin{table*}
\centering
 \contcaption{}
\begin{tabular}{lccccccc}
\hline 
\hline
Star  &  $\Pi^{\rm obs}$ & $A$          &  $\Pi^{\rm th}$  &  $\ell$ &  $k$ & $|\Delta|$  & $\Phi$ \\
      &  $[$s$]$        & $[$mma$]$  &  $[$s$]$     &         &      &  $[$s$]$  &   $[$s$]$ \\ 
\hline
\hline 
G38$-$29 $^{(3)}$  &    413.307 &    3.07  &  413.985   &  2 &   16   & 0.678  &  1.515\\
               &    432.354 &    3.57  &  434.227   &  2 &   17   & 1.873  & \\
               &    546.960 &    6.97  &  545.442   &  2 &   22   & 1.519  & \\
               &    705.970 &   18.44  &  707.049   &  1 &   16   & 1.079  & \\
               &    840.390 &    5.19  &  839.307   &  1 &   20   & 1.083  & \\
               &    899.971 &   10.59  &  896.903   &  2 &   38   & 3.068  & \\
               &    922.567 &    5.94  &  921.066   &  2 &   39   & 1.591  & \\
               &    945.448 &   12.34  &  946.328   &  2 &   40   & 0.880  & \\
               &    962.007 &    8.09  &  962.277   &  1 &   23   & 0.270  & \\
               &    963.593 &    4.58  &  962.277   &  1 &   23   & 1.316  & \\
               &    989.719 &   10.04  &  993.0267  &  2 &   42   & 3.308  & \\
               &    1002.16 &    7.14  &  1003.878  &  1 &   24   & 1.718  & \\ 
               &    1016.15 &    5.79  &  1014.220  &  2 &   43   & 1.930  & \\
               &    1081.82 &    5.04  &  1082.720  &  1 &   26   & 0.900  & \\
\hline 
PG 1541$+$650 $^{(4)}$ &    689.00     & $\cdots$ &  688.891   &  1 & 11 & 0.109 & 0.270 \\
            &    757.00     & $\cdots$ &  757.047   &  1 &   12   & 0.047  & \\
            &    564.00     & $\cdots$ &  563.346   &  2 &   16   & 0.654  & \\
\hline 
G191$-$16 $^{(8)}$ &   510.00    & $\cdots$ &  509.983   &  1 &    9   & 0.017  & 0.931 \\ 
              &   600.00    & $\cdots$ &  598.812   &  1 &   11   & 1.188  & \\
              &   710.00    & $\cdots$ &  712.027   &  1 &   14   & 2.027  & \\ 
              &   893.00    & $\cdots$ &  893.495   &  1 &   18   & 0.495  & \\
\hline 
G185$-$32     &   215.74   &   1.93   &  215.739   &  1 &    2   & 0.001 & 1.691 \\
              &   266.17   &   0.46   &  269.253   &  2 &    7   & 3.083 \\   
              &   300.60   &   1.04   &  298.724   &  2 &    8   & 1.876 \\
              &   370.21   &   1.62   &  367.694   &  1 &    5   & 2.516 \\
              &   651.70   &   0.67   &  652.677   &  1 &   12   & 0.978 \\
\hline 
EC 14012$-$1446 &   398.90    &  12.1    &  403.823   &  1 &    7   & 4.923  & 2.541 \\
                &   530.10    &  16.7    &  524.782   &  1 &   10   & 5.318  & \\
                &   610.40    &  54.3    &  613.677   &  1 &   12   & 3.277  & \\ 
                &   678.60    &   7.6    &  675.620   &  1 &   14   & 2.980  & \\
                &   722.90    &  22.9    &  721.733   &  1 &   15   & 1.167  & \\
                &   769.10    &  51.7    &  769.121   &  1 &   16   & 0.042  & \\
                &   882.70    &   2.9    &  883.878   &  2 &   34   & 1.178  & \\
                &   937.20    &  11.0    &  934.485   &  2 &   36   & 2.715  & \\
                &  1217.40    &   7.5    & 1216.141   &  1 &   27   & 1.259  & \\
\hline 
EC23487$-$2424 &   804.50    &  19.3    &  806.160   &  1 &   19   & 1.660  & 2.297 \\
               &   868.20    &  12.8    &  863.294   &  1 &   21   & 4.906  & \\
               &   992.70    &  24.4    &  992.375   &  1 &   24   & 0.325  & \\
\hline 
GD 165         &   114.30   &$\cdots$  &  114.278   &  2 &    2   & 0.022  & 0.889 \\    
               &   120.36   &$\cdots$  &  119.195   &  1 &    1   & 0.445  & \\   
               &   192.68   &$\cdots$  &  192.102   &  1 &    2   & 0.578  & \\   
               &   249.90   &$\cdots$  &  252.412   &  1 &    3   & 2.512  & \\  
\hline 
R548        &   187.28   &   0.9    &  187.597   &  1 &    1   & 0.308  & 2.516 \\ 
            &   212.95   &   5.4    &  213.401   &  1 &    2   & 0.451 \\
            &   274.51   &   3.5    &  242.263   &  1 &    3   & 2.249 \\ 
            &   318.07   &   1.1    &  311.361   &  2 &    8   & 6.709 \\  
            &   333.64   &   1.3    &  336.504   &  2 &    9   & 2.864 \\
\hline 
HE 1258$+$0123 &   439.20    &   9.8    &  446.066   & 2  &   14   & 6.867  & 2.099 \\ 
            &   528.50   &   9.3    &  527.704   & 1  &    9   & 0.796  & \\
            &   628.00    &  15.2    &  627.326   & 2  &   21   & 0.679  & \\
            &   744.60    &  22.9    &  744.780   & 1  &   14   & 0.180  &  \\
            &   881.50    &  17.6    &  892.728   & 1  &   17   & 1.228  & \\
            &  1092.10    &  14.1    & 1094.947   & 1  &   22   & 2.847  & \\
\hline 
GD 154      &   402.60    &   0.3    &  404.998   & 1  &    5   & 2.398  & 0.903 \\
            &  1088.60    &   2.0    & 1088.860   & 1  &   20   & 0.260  & \\
            &  1186.50    &   2.4    & 1186.550   & 1  &   22   & 0.050  & \\  
\hline 
\label{ajustes}
\end{tabular}
\end{table*}

\begin{table*}
\centering
 \contcaption{}
\begin{tabular}{lccccccc}
\hline 
\hline
Star  &  $\Pi^{\rm obs}$ & $A$          &  $\Pi^{\rm th}$  &  $\ell$ &  $k$ & $|\Delta|$  & $\Phi$ \\
      &  $[$s$]$        & $[$mma$]$  &  $[$s$]$     &         &      &  $[$s$]$  &   $[$s$]$ \\ 
\hline
\hline 
GD 385      &  128.10     &   3.7    &  130.665   & 2  &    2   & 2.564  & 1.291\\
            &  256.00     &  11.2    &  255.983   & 1  &    3   & 0.017  \\
\hline 
HE 1429$-$037 &  450.10     &  10.2    &  449.474   & 1  &    6   & 0.626  & 1.378 \\
            &  826.40     &  18.3    &  829.489   & 1  &   14   & 3.089  &  \\
            &  969.00     &  12.7    &  968.924   & 1  &   17   & 0.076  & \\
            & 1084.90     &  16.3    & 1080.279   & 1  &   19   & 4.621  & \\
\hline 
HS 1249$+$0426 & 288.90     &   7.55   &  288.905    & 1  &    4  & 0.005  & 0.005 \\
\hline 
G238$-$53     &  206.00     & 9.0 &  205.987    & 1  &    2  & 0.013  & 0.013 \\
\hline 
HS 1625$+$1231 $^{(5)}$  &  248.90     &   7.8    &  250.127   & 2  &    7   & 1.227  & 3.020\\
            &  268.20     &  13.3    &  274.612   & 1  &    3   & 6.412  &  \\
            &  325.50     &  13.3    &  320.910   & 1  &    5   & 4.590  & \\
            &  353.00     &  10.7    &  351.912   & 2  &   11   & 1.088  & \\
            &  385.20     &  17.0    &  382.670   & 1  &    6   & 2.530  & \\
            &  425.80     &  13.9    &  461.967   & 1  &    7   & 6.167  & \\
            &  533.60     &  23.6    &  531.504   & 1  &    9   & 2.096  & \\
            &  862.90     &  48.9    &  862.949   & 1  &   17   & 0.049  & \\
\hline 
G29$-$38    &  218.70     &   1.5    &  217.321   & 2  &    4   & 1.379  & 2.841\\  
            &  283.90     &   4.8    &  282.919   & 2  &    6   & 0.981  & \\
            &  363.50     &   4.7    &  365.551   & 2  &    9   & 2.051  & \\
            &  400.50     &   9.1    &  406.814   & 2  &   10   & 6.314  & \\
            &  496.20     &   7.9    &  493.659   & 2  &   13   & 2.541  & \\
            &  614.40     &  32.8    &  616.059   & 1  &    9   & 1.659  & \\
            &  655.10     &   6.1    &  644.728   & 2  &   18   & 10.372 & \\
            &  770.80     &   5.1    &  770.809   & 2  &   22   & 0.008  & \\
            &  809.40     &  30.1    &  800.395   & 2  &   23   & 9.005  & \\
            &  859.60     &  24.6    &  858.978   & 2  &   25   & 0.622  & \\
            &  894.00     &  14.0    &  891.098   & 2  &   26   & 2.902  & \\
            & 1150.50     &   3.6    & 1152.052   & 2  &   34   & 1.552  & \\
            & 1185.60     &   3.4    & 1185.529   & 2  &   35   & 0.072  & \\
            & 1239.90     &   1.9    & 1240.220   & 2  &   37   & 0.320  & \\
\hline 
PG2303$+$243$^{(6)}$  &  394.4    &  7.3     & 393.826    & 2  &  9     & 0.574  & 0.788\\
            &  616.4      & 31.4     & 616.560    & 1  &  8     & 0.160  & \\
            &  863.8      &  7.4     & 862.711    & 2  & 24     & 1.089  & \\
            &  965.3      & 19.7     & 966.590    & 1  & 15     & 1.290  & \\
\hline 
MCT0145$-$2211 & 462.20     &  25      &   462.353  & 1  &    7   & 0.153  & 1.494 \\ 
            &  727.90     &  19      &   726.912  & 1  &   13   & 0.988  & \\ 
            &  823.20     &  15      &   826.663  & 1  &   15   & 3.463  & \\
\hline 
BPM 30551   &  606.80     &  11.5    &   607.055  & 1  &   12   & 0.255  & 0.175 \\       
            &  744.70     &  10.5    &   744.605  & 1  &   15   & 0.096  & \\
\hline 
GD 99       & 1311.00       &   5.0    &  1311.002  & 1  &   28   & 0.002  & 0.002 \\
\hline 
BPM 24754   &     643.70  & $\cdots$ &   643.330  &  2 &   21   & 0.370  &0.938\\
            &    1045.10  & $\cdots$ &  1045.204  &  1 &   20   & 0.994  & \\
            &    1234.10  & $\cdots$ &  1234.005  &  1 &   24   & 0.095  & \\
            &    1356.60  & $\cdots$ &  1358.891  &  2 &   47   & 2.291  & \\ 
\hline  
KUV 02464$+$3239 $^{(7)}$ &  619.30     &   4.0    &   618.963  & 2  &   17   & 0.322  &  1.640\\ 
            & 777.60      &  5.5     &   779.541  & 1  &   12   & 1.931  & \\
            &  829.70     &  11.6    &   829.913  & 2  &   24   & 1.229  & \\
            &  866.20     &   9.5    &   860.447  & 2  &   25   & 5.704  & \\  
            &  993.20     &  13.2    &   992.707  & 1  &   16   & 0.717  & \\
            & 1250.30     &   4.4    &  1250.374  & 1  &   21   & 0.121  & \\
\hline 
PG 1149$+$058 & 1023.50     &  10.5    &  1023.479  & 1  &   20   & 0.021  & 0.021 \\
\hline 
BPM 31594 $^{(8)}$ & 401.93 & $\cdots$ & 402.453 & 1  &    5    & 0.523  & 0.321\\
            &  617.28    & $\cdots$ &   617.162 & 1  &   10    & 0.118  & \\
\hline
\label{ajustes}
\end{tabular}
\end{table*}

\begin{table*}
\centering
\contcaption{}
\begin{tabular}{lccccccc}
\hline 
\hline
Star  &  $\Pi^{\rm obs}$ & $A$          &  $\Pi^{\rm th}$  &  $\ell$ &  $k$ & $|\Delta|$  & $\Phi$ \\
      &  $[$s$]$        & $[$mma$]$  &  $[$s$]$     &         &      &  $[$s$]$  &   $[$s$]$ \\ 
\hline
\hline 
KUV 11370$+$4222 &  257.20   &   5.3    &  259.369   &  1 &    3   & 2.169  & 0.897 \\
            &    292.20   &   2.5    &  291.687   &  1 &    4   & 0.513  & \\
            &    462.90   &   3.2    &  462.919   &  2 &   15   & 0.019  & \\
\hline 
HS 1824$-$6000 $^{(5)}$ & 294.30 & 8.84 &   289.395  & 1  &    3   & 5.005  & 2.085\\ 
            & 304.40      &   7.66   &   301.198  & 2  &    8   & 3.202  & \\ 
            & 329.60      &  13.56   &   329.587  & 1  &    4   & 0.013  & \\
            & 384.40      &   3.30   &   384.520  & 2  &   11   & 0.120  & \\
\hline 
KUV 08368$+$4023 & 618.00   &  16.0    &   618.823  & 1  &   11   & 0.823  & 0.429 \\
            & 494.50      &   5.5    &   494.464  & 2  &   16   & 0.036  & \\
\hline 
R808 $^{(3)}$   & 404.46    &   1.99   &  400.923   & 2  &   14   & 3.534  & 3.499\\
            & 511.27    &   4.49   &  514.497   & 1  &   10   & 3.231  & \\
            & 632.18    &   3.41   &  629.270   & 2  &   24   & 2.909  & \\ 
            & 745.12    &   3.97   &  747.750   & 1  &   16   & 2.630  & \\
            & 796.25    &   3.97   &  799.402   & 2  &   31   & 3.149  & \\
            & 842.71    &   2.81   &  844.484   & 2  &   33   & 1.777  & \\
            & 860.23    &   3.48   &  865.257   & 2  &   34   & 5.030  & \\        
            & 875.15    &   3.73   &  870.376   & 1  &   19   & 4.770  & \\
            & 911.53    &   3.19   &  913.952   & 1  &   20   & 2.418  & \\ 
            & 915.80    &   5.54   &  615.230   & 2  &   36   & 0.573  & \\ 
            & 952.39    &   3.36   &  945.909   & 1  &   21   & 6.483  & \\ 
            & 960.53    &   3.68   &  967.199   & 2  &   38   & 6.672  & \\
            & 1011.39    &   2.54   &  1013.941  & 2  &   40   & 2.551  & \\
            & 1040.07    &   3.34   &  1038.204  & 2  &   41   & 1.866  & \\ 
            & 1066.73    &   2.21   &  1066.513  & 1  &   24   & 0.217  & \\
            & 1091.09    &   2.36   &  1084.277  & 2  &   43   & 6.813  & \\
            & 1143.96    &   2.50   &  1148.820  & 1  &   26 & 4.860  & \\ 
\hline 
G255$-$2      &  685.00     &  44      &  685.022   & 1  &   13   & 0.225  & 0.120 \\
            &  830.00     &  38      &  830.218   & 1  &   16   & 0.218  & \\
\hline 
HLTau$-$76    &  382.47    &  16.47   &  386.470   & 1  &    6   & 4.001  & 2.189 \\ 
            &  449.12    &    6.7   &  447.284   & 2  &   14   & 1.836  & \\
            &  492.12    &    7.12  &  494.302   & 1  &    8   & 2.182  & \\
            &  540.95    &   28.45  &  540.790   & 2  &   18   & 0.160  & \\
            &  596.79    &   14.40  &  595.623   & 1  &   10   & 1.167  & \\
            &  664.21    &   14.94  &  663.649   & 1  &   12   & 0.561  & \\
            &  781.00    &    9.1   &  789.047   & 2  &   27   & 8.047  & \\
            &  799.10    &    5.91  &  799.328   & 1  &   15   & 0.228  & \\ 
            &  933.64    &    2.40  &  933.879   & 1  &   18   & 0.239  & \\
            &  976.64    &    6.46  &  977.320   & 2  &   34   & 0.680  & \\
            & 1064.91    &   11.30  & 1064.845   & 1  &   21   & 0.065  & \\ 
            & 1390.84    &    3.92  & 1389.281   & 1  &   28   & 1.559  & \\
\hline 
G232$-$38     &  741.60     &    1.9   &  741.121   & 1  &   14   & 0.479  & 2.155 \\
            &  984.00     &    2.2   &  983.679   & 1  &   19   & 0.321  &  \\
            & 1147.50     &    1.9   & 1153.164   & 1  &   23   & 5.664  & \\
\hline 
G30$-$20      & 1068.00       &   13.8   & 1068.028   & 1  &   20   & 0.028  & 0.028\\
\hline 
\label{ajustes}
\end{tabular}\\
{\footnotesize $^{(1)}$ Bognar \& Papar\'o (2010), 
$^{(2)}$ Yeates et al. (2005), 
$^{(3)}$ Bischoff-Kim (2009), 
$^{(4)}$ Vauclair et al. (2000), 
$^{(5)}$ Voss et al. (2006), 
$^{(6)}$ Pak{\v s}tien{\.e}  et al. (2011), 
$^{(7)}$ Bognar et al (2009), 
$^{(8)}$ Bradley (1995)}
\end{table*}

\subsubsection{Some particular cases}

Because the large number of  ZZ Ceti stars seismologically analysed in
this work, it  would be unpractical and tedious  to describe in detail
the procedure  we followed to arrive at  the asteroseismological model
for  each  star,  as  we  already  did  for  the  particular  case  of
G117$-$B15A.   Instead,  we  briefly  summarise below  a  few  details
related to the selection process  of the best-fit model for some cases
of  interest.  The  structural parameters  of  the asteroseismological
models for  the complete set of  ZZ Ceti stars analysed  in this study
are shown in Table \ref{sismologia}.

\begin{table*}
\caption{Structural parameters  of the asteroseismological  models for
  the  sample of  ZZ Ceti  stars analyzed  in this  paper.  The quoted
  uncertainties   are  the  internal   errors  of   our  asteroseismic
  procedure.}  
\centering 
\scalebox{0.9}[0.9]{
\hspace{-7mm}
\begin{tabular}{lccccccccc}
\hline\hline
Star   &   $\log g$    & $T_{\rm eff}$ & $M_*/M_{\odot}$ &  $M_{\rm H} / M_*$  &   $M_{\rm He} / M_*$   & $\log (L/L_{\odot})$ &  $\log(R/R_{\odot})$ & $X_{\rm C}$ & $X_{\rm O}$\\ 
       &               & $[$K$]$      &                  &                    &                       &                     &                     &            &            \\
\hline
\hline
HS 1531$+$7436   & $8.28\pm 0.06$ & $12\,496\pm 210$ & $0.770\pm 0.034$ & $(1.55\pm 0.23)\times 10^{-5}$  & $5.96\times 10^{-3}$ & $-2.616\pm 0.011$ & $-1.977\pm 0.011$ & $0.332$ & $0.655$ \\ 
GD 244           & $7.97\pm 0.04$ & $12\,422\pm 105$ & $0.593\pm 0.012$ & $(1.17\pm 0.36)\times 10^{-4}$  & $2.38\times 10^{-2}$ & $-2.433\pm 0.011$ & $-1.881\pm 0.011$ & $0.283$ & $0.704$ \\
G226$-$29        & $8.28\pm 0.06$ & $12\,270\pm 290$ & $0.770\pm 0.034$ & $(2.02\pm 0.31)\times 10^{-5}$  & $5.95\times 10^{-2}$ & $-2.647\pm 0.011$ & $-1.977\pm 0.011$ & $0.332$ & $0.655$ \\
HS 0507$+$0434B  & $8.10\pm 0.06$ & $12\,257\pm 135$ & $0.660\pm 0.023$ & $(5.68\pm 1.94)\times 10^{-5}$  & $1.21\times 10^{-2}$ & $-2.532\pm 0.021$ & $-1.918\pm 0.016$ & $0.258$ & $0.729$ \\
LP 133$-$144     & $8.03\pm 0.04$ & $12\,210\pm 180$ & $0.609\pm 0.012$ & $(1.10\pm 0.79)\times 10^{-6}$  & $2.45\times 10^{-2}$ & $-2.507\pm 0.010$ & $-1.903\pm 0.011$ & $0.264$ & $0.723$ \\
EC 11507$-$1519  & $8.17\pm 0.07$ & $12\,178\pm 230$ & $0.705\pm 0.033$ & $(3.59\pm 1.09)\times 10^{-5}$  & $7.63\times 10^{-3}$ & $-2.592\pm 0.021$ & $-1.943\pm 0.016$ & $0.326$ & $0.661$ \\
L19$-$2          & $8.17\pm 0.07$ & $12\,105\pm 360$ & $0.705\pm 0.033$ & $(3.59\pm 1.66)\times 10^{-5}$  & $7.63\times 10^{-3}$ & $-2.602\pm 0.021$ & $-1.943\pm 0.016$ & $0.326$ & $0.661$ \\
GD 66            & $8.01\pm 0.04$ & $12\,068\pm 125$ & $0.593\pm 0.012$ & $(4.65\pm 4.37)\times 10^{-7}$  & $2.39\times 10^{-2}$ & $-2.514\pm 0.010$ & $-1.896\pm 0.011$ & $0.213$ & $0.704$ \\
G132$-$12        & $7.96\pm 0.05$ & $12\,067\pm 180$ & $0.570\pm 0.012$ & $(1.97\pm 0.46)\times 10^{-6}$  & $3.49\times 10^{-2}$ & $-2.486\pm 0.017$ & $-1.882\pm 0.014$ & $0.301$ & $0.606$ \\
G207$-$9         & $8.40\pm 0.07$ & $12\,029\pm 130$ & $0.837\pm 0.034$ & $(4.32\pm 3.50)\times 10^{-7}$  & $3.19\times 10^{-3}$ & $-2.761\pm 0.020$ & $-2.017\pm 0.016$ & $0.346$ & $0.641$ \\
G117$-$B15A      & $8.00\pm 0.09$ & $11\,985\pm 200$ & $0.593\pm 0.007$ & $(1.25\pm 0.70)\times 10^{-6}$  & $2.39\times 10^{-2}$ & $-2.497\pm 0.030$ & $-1.882\pm 0.029$ & $0.283$ & $0.704$ \\
MCT 2148$-$2911  & $8.05\pm 0.04$ & $11\,851\pm 150$ & $0.632\pm 0.014$ & $(7.58\pm 1.79)\times 10^{-5}$  & $1.75\times 10^{-2}$ & $-2.561\pm 0.011$ & $-1.904\pm 0.011$ & $0.232$ & $0.755$ \\
G38$-$29         & $8.28\pm 0.06$ & $11\,818\pm 50$  & $0.770\pm 0.034$ & $(1.23\pm 0.76)\times 10^{-5}$  & $5.96\times 10^{-3}$ & $-2.716\pm 0.011$ & $-1.979\pm 0.010$ & $0.333$ & $0.655$ \\
PG 1541$+$650    & $8.04\pm 0.04$ & $11\,761\pm 60$  & $0.609\pm 0.012$ & $(1.56\pm 1.42)\times 10^{-9}$  & $2.46\times 10^{-2}$ & $-2.583\pm 0.010$ & $-1.908\pm 0.011$ & $0.264$ & $0.723$ \\
G191$-$16        & $8.06\pm 0.04$ & $11\,741\pm 90$  & $0.632\pm 0.014$ & $(1.39\pm 0.32)\times 10^{-5}$  & $1.76\times 10^{-2}$ & $-2.590\pm 0.010$ & $-1.910\pm 0.011$ & $0.232$ & $0.755$ \\
G185$-$32        & $8.12\pm 0.10$ & $11\,721\pm 370$ & $0.660\pm 0.023$ & $(4.46\pm 3.20)\times 10^{-7}$  & $1.22\times 10^{-2}$ & $-2.632\pm 0.051$ & $-1.930\pm 0.034$ & $0.258$ & $0.729$ \\
EC14012$-$1446   & $8.05\pm 0.04$ & $11\,709\pm 95$  & $0.632\pm 0.014$ & $(7.58\pm 2.40)\times 10^{-5}$  & $1.75\times 10^{-2}$ & $-2.583\pm 0.011$ & $-1.904\pm 0.011$ & $0.232$ & $0.755$ \\
EC23487$-$2424   & $8.28\pm 0.06$ & $11\,700\pm 75$  & $0.770\pm 0.034$ & $(2.02\pm 0.32)\times 10^{-5}$  & $5.95\times 10^{-3}$ & $-2.731\pm 0.010$ & $-1.978\pm 0.010$ & $0.332$ & $0.655$ \\
GD 165	         & $8.05\pm 0.07$ & $11\,635\pm 330$ & $0.632\pm 0.014$ & $(7.58\pm 3.28)\times 10^{-5}$  & $1.75\times 10^{-2}$ & $-2.594\pm 0.043$ & $-1.904\pm 0.029$ & $0.232$ & $0.755$ \\
R548             & $8.03\pm 0.05$ & $11\,627\pm 390$ & $0.609\pm 0.012$ & $(1.10\pm 0.38)\times 10^{-6}$  & $2.45\times 10^{-2}$ & $-2.594\pm 0.025$ & $-1.904\pm 0.015$ & $0.264$ & $0.723$ \\
HE 1258$+$0123   & $8.07\pm 0.03$ & $11\,582\pm 100$ & $0.632\pm 0.014$ & $(4.46\pm 3.07)\times 10^{-6}$  & $1.76\times 10^{-2}$ & $-2.620\pm 0.014$ & $-1.913\pm 0.007$ & $0.232$ & $0.755$ \\
GD 154           & $8.20\pm 0.04$ & $11\,574\pm 30$  & $0.705\pm 0.033$ & $(4.58\pm 1.80)\times 10^{-10}$ & $7.66\times 10^{-3}$ & $-2.705\pm 0.003$ & $-1.955\pm 0.003$ & $0.326$ & $0.661$ \\
GD 385           & $8.07\pm 0.03$ & $11\,570\pm 90$  & $0.632\pm 0.014$ & $(4.59\pm 2.86)\times 10^{-7}$  & $1.76\times 10^{-2}$ & $-2.628\pm 0.005$ & $-1.962\pm 0.005$ & $0.232$ & $0.755$ \\
HE 1429$-$037    & $8.13\pm 0.05$ & $11\,535\pm 85$  & $0.660\pm 0.023$ & $(4.68\pm 0.86)\times 10^{-10}$ & $1.22\times 10^{-3}$ & $-2.667\pm 0.018$ & $-1.934\pm 0.013$ & $0.258$ & $0.729$ \\
HS 1249$+$0426   & $8.02\pm 0.02$ & $11\,521\pm 35$  & $0.609\pm 0.012$ & $(3.53\pm 1.08)\times 10^{-5}$  & $2.45\times 10^{-2}$ & $-2.595\pm 0.002$ & $-1.896\pm 0.002$ & $0.264$ & $0.723$ \\
G238$-$53        & $8.03\pm 0.02$ & $11\,497\pm 120$ & $0.609\pm 0.012$ & $(1.54\pm 0.28)\times 10^{-6}$  & $2.46\times 10^{-2}$ & $-2.613\pm 0.002$ & $-1.904\pm 0.002$ & $0.264$ & $0.723$ \\
HS 1625$+$1231   & $8.02\pm 0.04$ & $11\,485\pm 230$ & $0.609\pm 0.012$ & $(3.52\pm 1.67)\times 10^{-5}$  & $2.45\times 10^{-2}$ & $-2.600\pm 0.016$ & $-1.896\pm 0.012$ & $0.264$ & $0.723$ \\
G29$-$38         & $8.01\pm 0.03$ & $11\,471\pm 60$  & $0.593\pm 0.012$ & $(4.67\pm 2.83)\times 10^{-10}$ & $2.39\times 10^{-2}$ & $-2.612\pm 0.006$ & $-1.901\pm 0.006$ & $0.283$ & $0.704$ \\
PG2303$+$242     & $7.88\pm 0.07$ & $11\,210\pm 100$ & $0.525\pm 0.12 $ & $(4.54\pm 2.95)\times 10^{-8}$  & $4.94\times 10^{-2}$ & $-2.579\pm 0.03$  & $-1.865\pm 0.032$ & $0.279$ & $0.709$ \\
MCT 0145$-$2211  & $7.95\pm 0.03$ & $11\,439\pm 120$ & $0.570\pm 0.012$ & $(1.43\pm 0.38)\times 10^{-5}$  & $3.50\times 10^{-2}$ & $-2.573\pm 0.014$ & $-1.879\pm 0.012$ & $0.301$ & $0.686$ \\
BPM 30551        & $8.19\pm 0.05$ & $11\,435\pm 40$  & $0.705\pm 0.033$ & $(4.36\pm 0.26)\times 10^{-6}$  & $7.66\times 10^{-3}$ & $-2.714\pm 0.006$ & $-1.949\pm 0.006$ & $0.326$ & $0.661$ \\
GD 99            & $8.01\pm 0.13$ & $11\,395\pm 25$  & $0.660\pm 0.023$ & $(1.36\pm 0.52)\times 10^{-5}$  & $1.22\times 10^{-2}$ & $-2.671\pm 0.005$ & $-1.950\pm 0.068$ & $0.258$ & $0.729$ \\  
BPM 24754        & $8.03\pm 0.03$ & $11\,390\pm 50$  & $0.609\pm 0.012$ & $(4.51\pm 2.72)\times 10^{-6}$  & $2.46\times 10^{-2}$ & $-2.626\pm 0.011$ & $-1.902\pm 0.001$ & $0.264$ & $0.723$ \\
KUV 02464$+$3239 & $7.93\pm 0.03$ & $11\,360\pm 40$  & $0.548\pm 0.014$ & $(4.71\pm 2.45)\times 10^{-8}$  & $4.21\times 10^{-2}$ & $-2.579\pm 0.006$ & $-1.876\pm 0.006$ & $0.290$ & $0.697$ \\
PG 1149$+$058    & $7.94\pm 0.02$ & $11\,336\pm 20$  & $0.570\pm 0.012$ & $(5.29\pm 2.45)\times 10^{-5}$  & $3.69\times 10^{-2}$ & $-2.579\pm 0.001$ & $-1.875\pm 0.002$ & $0.301$ & $0.686$ \\
BPM 31594        & $7.86\pm 0.03$ & $11\,250\pm 70$  & $0.525\pm 0.012$ & $(5.36\pm 1.87)\times 10^{-5}$  & $4.93\times 10^{-2}$ & $-2.545\pm 0.009$ & $-1.851\pm 0.009$ & $0.279$ & $0.709$ \\
KUV 11370$+$4222 & $8.06\pm 0.03$ & $11\,237\pm 80$  & $0.632\pm 0.014$ & $(1.40\pm 0.64)\times 10^{-5}$  & $1.76\times 10^{-2}$ & $-2.668\pm 0.007$ & $-1.911\pm 0.008$ & $0.232$ & $0.755$ \\  
HS 1824$-$6000   & $7.95\pm 0.08$ & $11\,234\pm 400$ & $0.570\pm 0.012$ & $(1.43\pm 0.62)\times 10^{-5}$  & $3.50\times 10^{-2}$ & $-2.605\pm 0.050$ & $-1.879\pm 0.030$ & $0.301$ & $0.686$ \\
KUV 08368$+$4026 & $8.02\pm 0.03$ & $11\,230\pm 95$  & $0.609\pm 0.012$ & $(1.42\pm 0.52)\times 10^{-5}$  & $2.45\times 10^{-2}$ & $-2.646\pm 0.010$ & $-1.899\pm 0.007$ & $0.264$ & $0.723$ \\
R808             & $8.18\pm 0.05$ & $11\,213\pm 130$ & $0.705\pm 0.033$ & $(3.59\pm 1.70)\times 10^{-5}$  & $7.63\times 10^{-3}$ & $-2.738\pm 0.008$ & $-1.944\pm 0.008$ & $0.326$ & $0.661$ \\
G255$-$2         & $8.11\pm 0.04$ & $11\,185\pm 30$  & $0.660\pm 0.023$ & $(4.45\pm 2.12)\times 10^{-6}$  & $1.22\times 10^{-2}$ & $-2.709\pm 0.002$ & $-1.928\pm 0.003$ & $0.258$ & $0.729$ \\ 
HLTau$-$76       & $7.89\pm 0.03$ & $11\,111\pm 50$  & $0.548\pm 0.012$ & $(1.83\pm 1.03)\times 10^{-4}$  & $4.19\times 10^{-2}$ & $-2.579\pm 0.005$ & $-1.857\pm 0.005$ & $0.323$ & $0.697$ \\
G232$-$38        & $7.99\pm 0.04$ & $10\,952\pm 120$ & $0.593\pm 0.012$ & $(5.19\pm 1.87)\times 10^{-5}$  & $2.38\times 10^{-2}$ & $-2.666\pm 0.015$ & $-1.888\pm 0.010$ & $0.283$ & $0.704$ \\
G30$-$20         & $7.91\pm 0.02$ & $10\,950\pm 15$  & $0.548\pm 0.012$ & $(5.34\pm 2.18)\times 10^{-5}$  & $4.20\times 10^{-2}$ & $-2.618\pm 0.002$ & $-1.863\pm 0.002$ & $0.290$ & $0.697$ \\
\hline
\hline
\label{sismologia}
\end{tabular}
}
\end{table*}

\begin{itemize}

\item[-] {\bf G226$-$29.}  G226$-$29  also exhibits a single mode with
  a short period.  Fortunately, there exist a robust constraint on its
  $\ell$-identification. In fact, Kepler et al. (2005b) found that the
  mode is actually a triplet ($\ell= 1$) with the central component at
  a period  of 109.278 s. The  solution in this case  corresponds to a
  rather  massive  model with  a  thick  H  envelope ($M_*  \sim  0.77
  M_{\odot}$, $M_{\rm H} = 2.02 \times 10^{-5} M_*$), in line with the
  spectroscopic observations.

\item[-] {\bf HS 1531$+$7436.} This star exhibits a single mode with a
  very   short   period   (for   ZZ   Ceti   standards)   at   $112.5$
  s.  Unfortunately,  the  presence  of  just one  period  turns  very
  difficult any attempt  of asteroseismology on this star,  and we are
  forced to  make a somewhat  arbitrary assumption. If we  assume that
  this mode corresponds to a  $(\ell, k)= (1, 1)$ identification, then
  the  stellar mass  of the  seismological model  must be  larger than
  $0.705  M_{\odot}$.  In  the first  attempt to  fit its  periods, we
  obtained  massive  solutions ($M_*  \sim  0.77  M_{\odot}$), but  at
  effective  temperatures excessively low  ($\sim 10\,800$  K).  These
  solutions  are characterized by  thick H  envelopes.  Since  we have
  just a single  observed period, it is possible to  find a model with
  the appropriate H  envelope thickness as to allow  to fit the period
  at   an  effective   temperature   in  close   agreement  with   the
  spectroscopic value of $T_{\rm eff}$.   To this end, we selected the
  sequence  with $M_*  = 0.77  M_{\odot}$ and  computed  an additional
  sequence with $M_{\rm H} = 1.55 \times 10^{-5} M_*$. In this way, we
  obtained a best-fit model with $T_{\rm eff} \simeq 12\, 350$ K.

\item[-] {\bf G185$-$32.} The  pulsation spectrum of this DAV includes
  a  period at  215.74 s,  quite  similar to  the dominant  mode in  
  G117$-$B15A,  but  at variance  with  this  star,  the difference  of
  amplitude between this mode and  the remaining ones is not so strong
  in the case  of G185$-$32.  The identification of  the $\ell$ degree
  for the periodicities observed in G185$-$32 is not well determined. In
  particular, the period at $\sim  215$ s is associated with a $\ell=
  1$ or $\ell=2$ mode (Castanheira et al.  2004; Yeates et al.  2005).
  Similarly to G117$-$B15A,  for this star the stellar  models fit the
  period at  215.74 s with a  mode characterized by $\ell=  1$ and $k=
  2$. However, the  seismological model for this star  is more massive
  than  in  the case  of  G117$-$B15A.  For  G185$-$32 we  adopted  an
  asteroseismological model that closely  fit the period at $\sim 215$
  and at  the same time  it matches the  set of observed  periods with
  mostly $\ell= 1$ modes.

\item[-] {\bf GD 154.} This star shows three pulsation modes. The mode
  with period  at 402.6  s is  a unstable and  low amplitude  mode, as
  compared with the remaining two modes (Pfeiffer et al.  1996). Since
  the amplitude  of the  long period modes  (1088.6 and 1186.5  s) are
  very similar, and since the period  at 1186.5 s is probably a dipole
  mode (Pfeiffer et al.  1996)  we favor models that fit these periods
  with $\ell= 1$ modes.  Generally,  the solutions have a stellar mass
  between  $0.6323$  and  $0.705  M_{\odot}$  with  thin  H  envelopes
  ($M_{\rm H}  \sim 10^{-8}-10^{-10} M_*$). Among them,  we choose the
  solution with $M_*  = 0.705 M_{\odot}$ and $M_{\rm  H} = 4.58 \times
  10^{-10}M_*$  because it  has surface  parameters in  agreement with
  spectroscopy.   Other similar solutions  have $T_{\rm  eff}\sim 11\,
  200$ K, but in these cases  the period at 402.6 s is identified with
  $\ell =2$.

\item[-]  {\bf G238$-$53,  G132$-$12} and  {\bf LP  133$-$144.}  These
  three stars  also have a period near  215 s. In all  the cases, this
  mode has an identification $k= 2$ when $\ell= 1$. This $\ell$ and $k$
  identification  is   an  intrinsic   property  shared  by   all  the
  asteroseismological  models of  this study.  This can  be  seen from
  Fig. \ref{todos}, that shows that the periods with $\ell= 1$ and $k=
  1$ are always too short to match the $\sim 215$ s period.

\item[-]  {\bf R548.}  This  star  has a  slightly higher  effective
  temperature and  a spectroscopic  stellar mass a  bit larger  than 
  G117$-$B15A.   Frequently, both  stars are  analyzed together  due to
  these  similarities and  several periods  in common.   In  the first
  attempts to fit the periods of R548 we obtained solutions with high
  mass, but  were discarded because  the 212 s periods  was identified
  with $\ell=  2$ according to  those models. Also,  intermediate mass
  solutions were obtained. Generally, the modes with periods at 318.07
  s and 333.64 s are the most poorly matched by the models, and they are
  identified with  $\ell= 2$. In order  to found a  best-fit model for
  this star, we were forced to employ several restrictions. We assumed
  that the mode with the period at 212.95 s has $\ell=1$ {\bf and} 
  $k =2$, and
  fixed  $\ell= 1$  also for  the mode  with the  period at  274.272 s
  (Yeates et  al. 2005). We found an  asteroseismological model with
  $M_*= 0.609 M_{\odot}$, larger than  the stellar mass obtained for 
  G117$-$B15A ($M_*=  0.593 M_{\odot}$) and  a $T_{\rm eff}$  lower, in
  contrast to the trend indicated  by spectroscopy and by the previous
  studies (Bradley  1998; Castanheira  \& Kepler 2009).   However, the
  surface parameters characterizing the  best fit model are within the
  uncertainties of spectroscopy.

\item[-] {\bf MCT 2148$-$2911,  PG 1541$+$650, HE 1429$-$037} and {\bf
  HS 1824$-$600.}  These are low-mass white dwarfs, with spectroscopic
  masses of $0.515, 0.502, 0.514$ and $0.427 M_{\odot}$, respectively.
  These  values are  obtained by  extrapolation from  our evolutionary
  model grid.  However, our asteroseismological models for these stars
  do not have  the lowest mass of our  model grid ($0.525 M_{\odot}$),
  but  instead, they  result  in intermediate  masses: $0.632,  0.609,
  0.660$ y $0.570 M_{\odot}$,  respectively.  The $T_{\rm eff}$ values
  of these models are in agreement with the spectroscopic inferences.

\item[-] {\bf  GD 244.} For this star  we have not been  able to found
  any  plausible  seismological model  with  an effective  temperature
  close to  the spectroscopic value  ($T_{\rm eff}= 11\, 680$  K).  In
  order to adopt  a seismological model, we considered  that the large
  amplitude modes are $\ell= 1$, and found an acceptable solution with
  a $T_{\rm  eff}= 12\,422$ K, markedly higher  than the spectroscopic
  one. On the other hand, the  gravity and stellar mass of the adopted
  seismological model  are compatible (within  the uncertainties) with
  the spectroscopic estimates.

\item[-]  {\bf   G207$-$9.}   For  this   DAV  we  obtain   a  massive
  seismological  solution, with  $M_*= 0.837  M_{\odot}$.   However, a
  second  solution,  although  with  a slightly  worse  period  match,
  according to  $\Phi= 1.496$ s, is  obtained for a  lower mass ($M_*=
  0.609 M_{\odot}$), characterized by  a thick H envelope ($M_{\rm H}=
  1.41 \times 10^{-5}  M_*$). A degeneracy of solutions  for this star
  has been also found by Castanheira \& Kepler (2009).

\item[-] {\bf G29$-$38.} This is a rather pathological case.  In spite
  of the  fact that this  star has $T_{\rm  eff} \sim 11\, 800$  K, it
  exhibits a  rich and complex  period spectrum (including  14 genuine
  eigenmodes)  which is  characteristic of  cooler DAVs.   Thompson et
  al.  (2008), by means  of VLT  spectroscopy, show  that most  of the
  periodicities exhibited by this star  are $\ell= 1$ modes, but there
  are also some  $\ell= 2$ modes and possibly one  mode with $\ell= 3$
  or  $4$. However, the  seismological solutions  for this  star imply
  that  most  of  the observed  modes  should  be  $\ell= 2$.  In  the
  asteroseismological  model  adopted,  the  only $\ell=  1$  mode  is
  associated to the  mode with a period 614.4 s  which has the largest
  amplitude.

\item[-] {\bf PG  2303$-$243.}  For this star, the  observed modes and
  amplitudes were taken from  Pak{\v s}tien{\.e} et al. (2011).  These
  authors show  that this ZZ Ceti  has a very  rich pulsation spectrum
  with $24$  probably independent modes. However, most  of these modes
  show very  low amplitudes, below $\sim  4$ mma. In  our analysis, we
  considered  to be real  modes only  those showing  amplitudes higher
  than  $\sim 4$  mma, leaving  us with  just four  periodicities.  In
  particular, we fixed the harmonic degree to be $\ell= 1$ for the two
  main modes, 616.4 and 965.3 s, while we allow the remainder modes to
  be $\ell= 1$ or $2$.

\end{itemize}

\subsubsection{Seismic stellar masses}

\begin{figure}
\begin{center}
\includegraphics[clip,width=240pt]{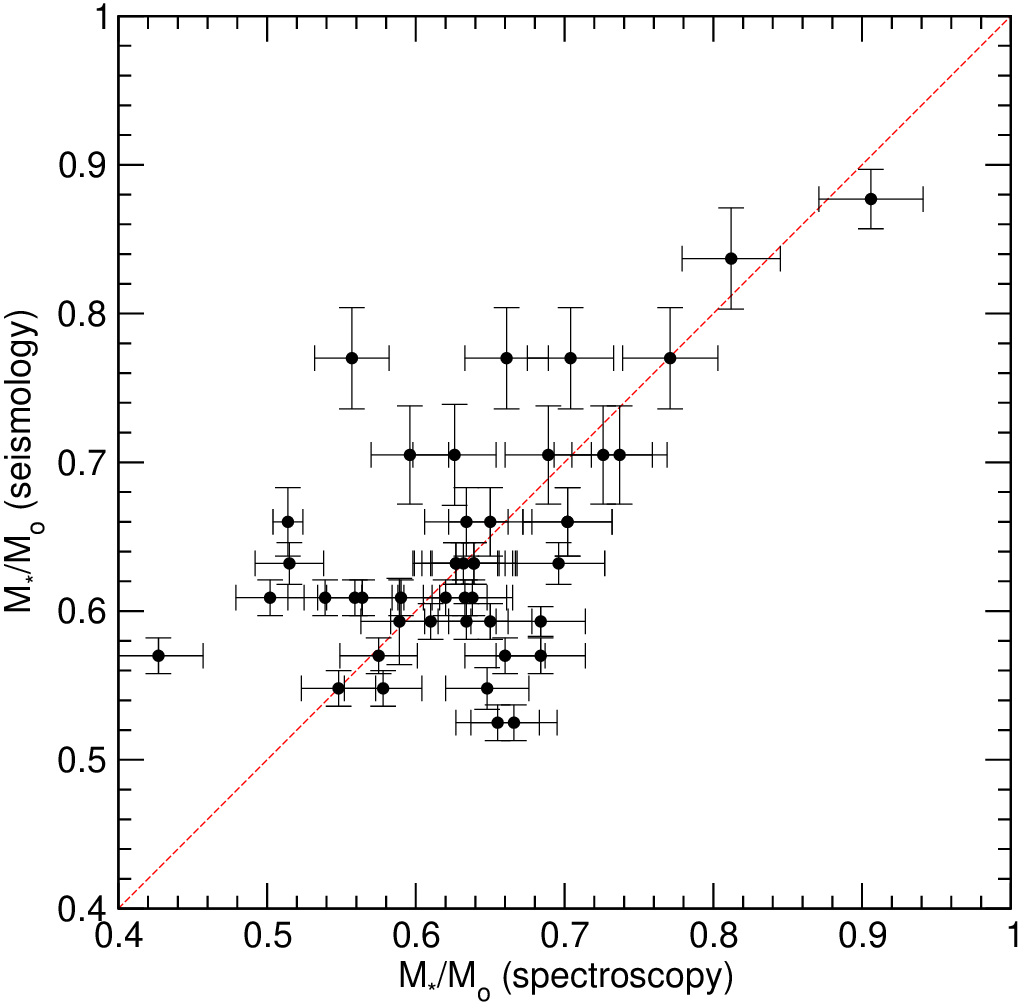}
\caption{Comparison between  the value of  the stellar mass of  the 44
  DAVs  stars analyzed in  this work,  according to  our spectroscopic
  inference  (x$-$axis)  and  from  our  asteroseismological  analysis
  (y$-$axis). The  red dashed line represents a  perfect match between
  both mass estimates.}
\label{massmass}
\end{center}
\end{figure}

\begin{figure}
\begin{center}
\includegraphics[clip,width=240pt]{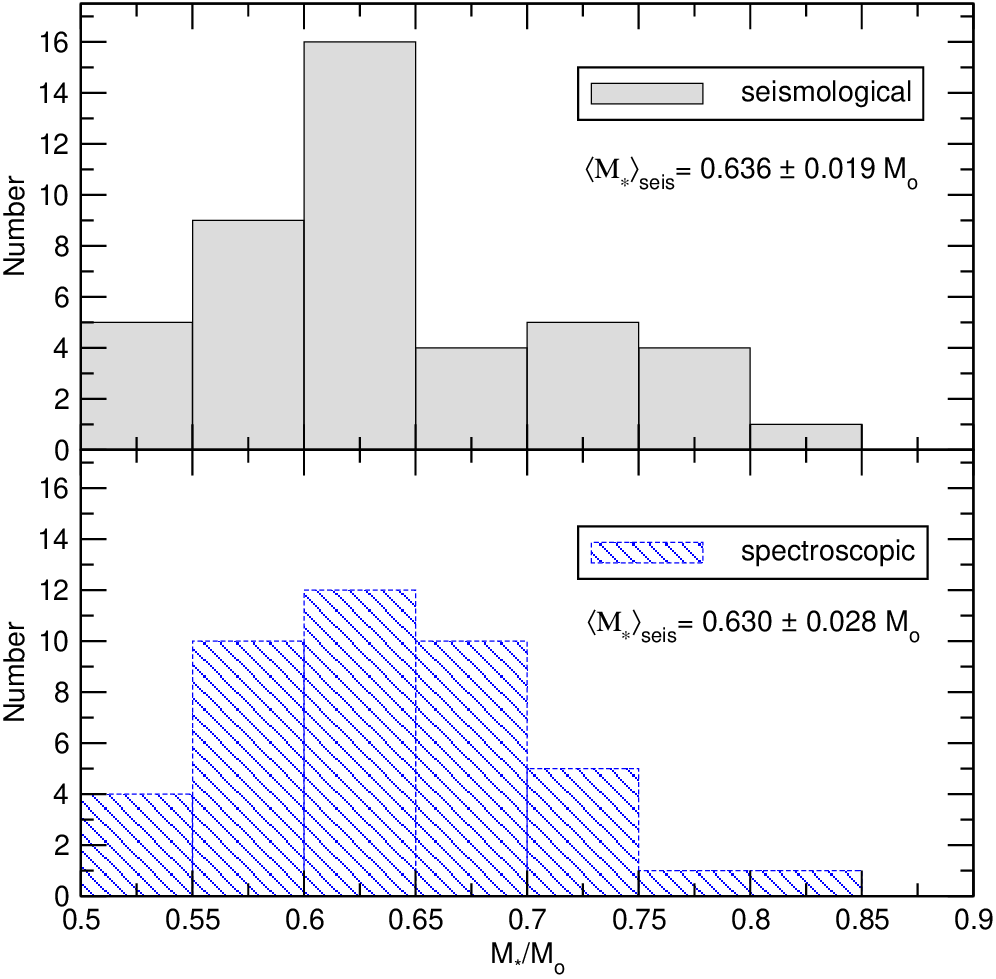}
\caption{Histograms showing the mass distribution for the sample of 44
  ZZ   Ceti  stars  considered   in  this   work,  according   to  our
  spectroscopic  inferences   (lower  panel)  and   our  seismological
  analysis (upper panel).}
\label{histo-mass}
\end{center}
\end{figure}

In this  work, the DA white  dwarf evolutionary tracks  used to derive
the spectroscopic masses  of the DAVs have been  employed to infer the
asteroseismological masses.   Thus, a comparison between  both sets of
values  is  worth  doing.   We  compare in  Fig.   \ref{massmass}  the
spectroscopic and asteroseismological masses.   The dotted line is the
1:1  correspondence.   The plot  reveals  that  the general  agreement
between both  sets of estimations is  far from being  good, the larger
discrepancies  reaching  differences   up  to  $\sim  0.2  M_{\odot}$.
However,  the bulk  of the  points in  Fig.  \ref{massmass} accumulate
around  the  dotted line,  demonstrating  that  no appreciable  offset
exists between the spectroscopic  and asteroseismic estimations of the
stellar mass.

The distribution  of stellar masses according  to asteroseismology and
spectroscopy  is depicted  in the  histograms of  the upper  and lower
panel of Fig.  \ref{histo-mass},  respectively.  The mean value of the
asteroseismological mass  is $\langle M_*\rangle_{\rm  seis}= 0.636\pm
0.019  M_{\odot}$,   slightly  larger   ($\sim  0.95  \%$)   than  the
spectroscopic  one,  $\langle  M_*\rangle_{\rm spec}=  0.630\pm  0.028
M_{\odot}$\footnote{We do not claim the pulsators are more massive, as
  there are  strong selection effects  in the search  for pulsators.}.
Given the  very different methods  employed to infer both  values, the
excellent agreement between these average masses is encouraging.

Castanheira  \&   Kepler  (2008,   2009)  have  performed   the  first
asteroseismological study of an ensemble  of ZZ Ceti stars.  They have
studied a total of 83 ZZ Ceti stars including the bright variables and
also a subset  of the SDSS variables. The average mass  of the ZZ Ceti
stars as  derived by these authors is  $\langle M_*\rangle_{\rm seis}=
0.668  M_{\odot}$,  about  $5  \%$  higher than  our  value,  $\langle
M_*\rangle_{\rm seis}=  0.636 M_{\odot}$. We note  that Castanheira \&
Kepler (2008, 2009)  have included several very massive  ZZ Ceti stars
($M_*  \gtrsim 1  M_{\odot}$) that  have  not been  considered in  our
study. Given  the fact that the  numerical tools used  in modeling the
structure, evolution and  pulsations of ZZ Ceti stars  used by the two
groups are independent,  and given that the samples  of stars analysed
are not the same, we consider that the $\langle M_*\rangle_{\rm seis}$
value derived in  this work and that derived  by Castanheira \& Kepler
(2008, 2009) are in very good agreement.

\subsubsection{The thicknesses of the hydrogen envelope}

\begin{figure}
\begin{center}
\includegraphics[clip,width=240pt]{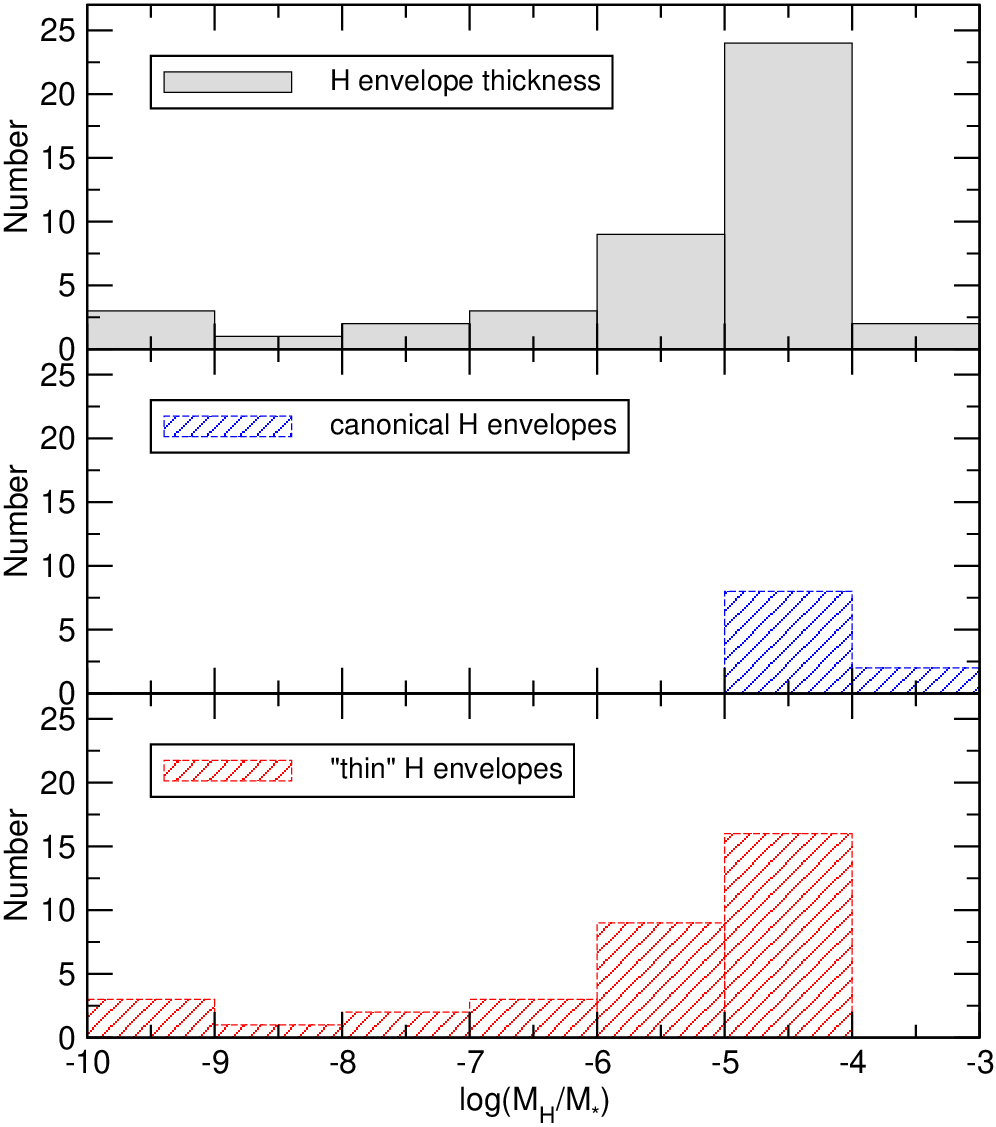}
\caption{Upper  panel:  histogram  showing  the H  envelope  thickness
  distribution for the  sample of 44 ZZ Ceti  stars considered in this
  work.  Middle  panel: histogram for models with  canonical (thick) H
  envelope  thicknesses, as predicted by canonical
evolutionary computations according to the value of the stellar mass.  
Lower   panel:  histogram  for  models  with
  non-canonical (thin) envelope thicknesses, as obtained by means of 
our artificial procedure described in Sect. \ref{grid}.}
\label{histo-mh}
\end{center}
\end{figure}

\begin{figure}
\begin{center}
\includegraphics[clip,width=240pt]{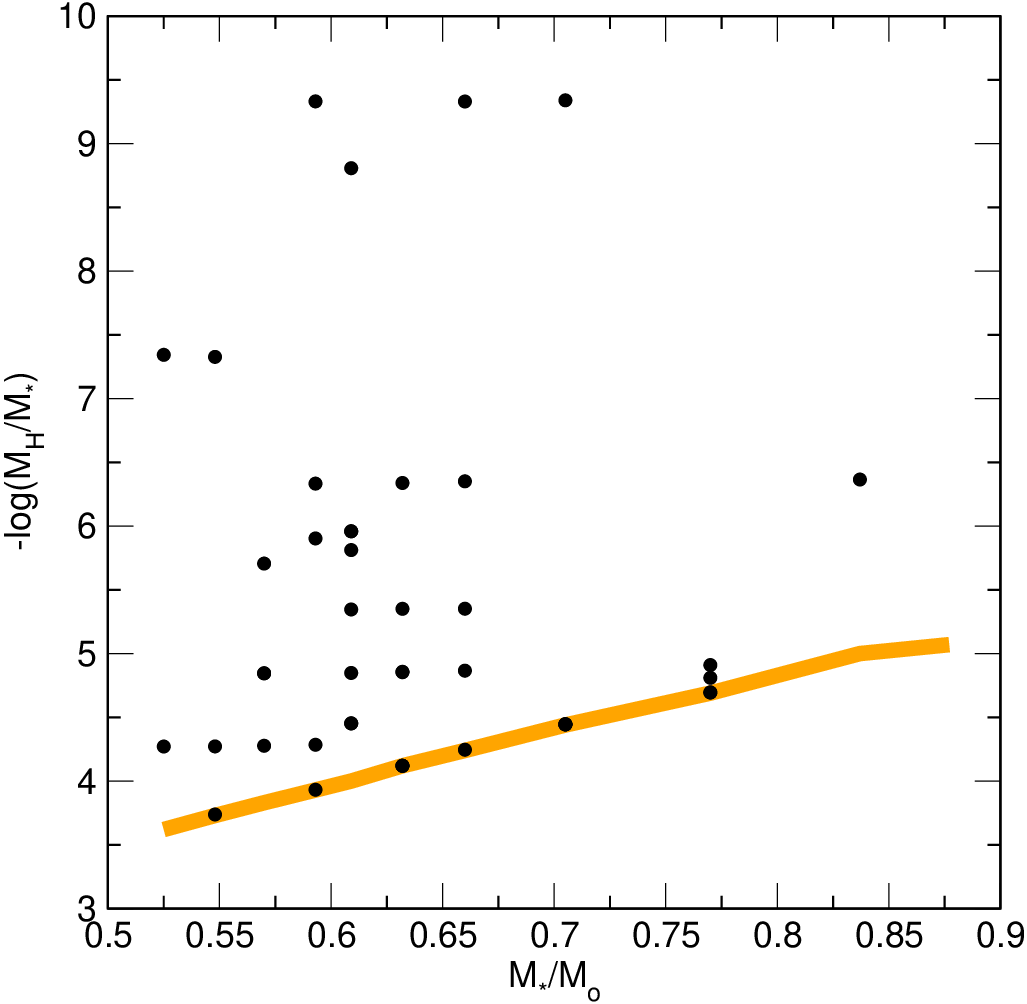}
\caption{The values  of the  H envelope mass  versus the  stellar mass
  corresponding to  the asteroseismological models  of the 44  ZZ Ceti
  stars analysed  in this work.   The thick (orange) curve  depicts the
  canonical values for the H envelope thickness.}
\label{mh-m-all}
\end{center}
\end{figure}

One of the  most important structural parameters we  want to constrain
through asteroseismology  of ZZ Ceti stars  is the thickness  of the H
envelope in DA white dwarfs.  We  have found a H layer mass of $M_{\rm
  H}=  (1.25\pm 0.7) \times  10^{-6} M_*$  for G117$-$B15A,  about two
order  of magnitude  thinner  than the  value  predicted by  canonical
evolutionary computations, of $M_{\rm  H} \sim 10^{-4} M_*$. Here, the
analysis of a  large number of ZZ Ceti stars allows  us to explore the
distribution of H envelope thicknesses from their pulsations.  In Fig.
\ref{histo-mh} we present histograms of the distribution of H envelope
thicknesses.  In the upper panel  we show the results for the complete
sample of  44 stars. Note  that there is  a pronounced maximum  of the
distribution for  $\log (M_{\rm  H}/M_*)$ in the  range $-5$  to $-4$,
although there  exists another, much less notorious  maximum for $\log
(M_{\rm H}/M_*)$ between  $-10$ and $-9$. So, it  is apparent from the
figure  that  there exists  a  {\it range}  of  thicknesses  of the  H
envelope  in  the studied  DAV  stars, with  a  strong  peak at  thick
envelopes and another  much lower peak at very  thin envelopes, and an
apparent paucity for intermediate thicknesses.  In the middle panel of
Fig.   \ref{histo-mh}  we  show  the histogram  corresponding  to  the
asteroseismological  models  characterized   by  canonical  (thick)  H
envelope thicknesses, that  amount to 11 stars. Finally,  in the lower
panel we display the histogram for the non-canonical thicknesses, that
is, envelopes  thinner than  those predicted by  standard evolutionary
computations  depending on  the  value  of the  stellar  mass.  As  in
previous  sections,  we  refer  this  kind of  envelopes  as  ``thin''
envelopes.   We  recall  that   these  ``thin''  envelopes  have  been
generated  in this  work in  order to  extend the  exploration  of the
parameter space of the models for asteroseismology.  Note that in most
of  the   analysed  stars   (34  stars  from   a  total  of   44)  our
asteroseismological models  have ``thin'' H  envelopes, as illustrated
in Fig. \ref{mh-m-all}. It is important to note, however, that most of
our  derived envelope masses,  even being  thinner than  the canonical
values,  cluster close to  the envelope  masses predicted  by standard
evolutionary computations,  at variance  with those of  Castanheria \&
Kepler (2009), who found a nearly homogeneous distribution of envelope
masses in their fits (see their Fig. 8).

The mean value of the H  layer mass is $\langle M_{\rm H}/M_* \rangle=
2.71 \times 10^{-5}$ according to  our results. This value is about 50
times larger than  the value obtained by Castanheira  \& Kepler (2009)
with different  samples, $\langle  M_{\rm H}/M_* \rangle=  5.01 \times
10^{-7}$.  In  spite of  this difference, both  studies concur  to the
conclusion that  an important fraction  of DA white dwarfs  might have
been formed with a H mass smaller than the value predicted by standard
evolutionary computations,  a conclusion we have  already suggested at
end  of  Section \ref{g117b15a}  on  the basis  of  our  results on  
G117$-$B15A.

\section{Conclusions}
\label{conclusions}

In  this paper,  we  have carried  out  the first  asteroseismological
application  of the  evolutionary DA  white-dwarf models  presented in
Althaus et  al. (2010b)\footnote{Detailed tabulations  of the chemical
  profiles for different stellar masses and effective temperatures are
  available at  our web site http://www.fcaglp.unlp.edu.ar/evolgroup}.
Specifically, we performed a  detailed asteroseismological study of 44
ZZ Ceti  stars extracted from a  sample of bright stars  for which the
surface  parameters are  accurately  known. This  sample includes  the
archetypal ZZ Ceti star G117$-$B15A. The asteroseismological analysis 
of such a large set of stars has the potential to characterize the 
common properties of the class. We have employed a large grid of
fully  evolutionary   models  characterized  by   consistent  chemical
profiles from the  centre to the surface and covering  a wide range of
stellar  masses,   thicknesses  of   the  H  envelope   and  effective
temperatures.    Our   asteroseismological   approach   represents   a
significant improvement  over previous  calculations that rely  on the
use  of DA  white dwarf  models characterized  by  simplified chemical
profiles  at the envelope  and/or the  core. This  is the  first work
aimed at an asteroseismological analysis of ZZ Ceti stars that employs
{\sl fully evolutionary} white dwarf models.

Our main results for G117$-$B15A are:

\begin{itemize}

\item  We  found an  asteroseismological  model  for G117$-$B15A  with
  $T_{\rm eff}= 11\,985  \pm 200$ K, $\log g=  8.00\pm 0.09$ and $M_*=
  0.593  \pm  0.007  M_{\odot}$,   in  excellent  agreement  with  the
  spectroscopic determinations.

\item  For   the  first   time,  we  break   the  degeneracy   of  the
  asteroseismological  solutions for  this star  reported  by previous
  studies regarding the thickness of  the H envelope, depending on the
  $k$-identification  of the three  periods exhibited  by G117$-$B15A,
  although it is fair to say that we are matching 3 periods by varying
  3 parameters.  We found the  identification $k= 2, 3, 4$ as the
  only possible one in the frame of our set of pulsation models.

\item Our  best-fit model has a  H envelope with  $M_{\rm H}= (1.25\pm
  0.7) \times 10^{-6} M_*$, about  two order of magnitude thinner than
  the  value  predicted  by  canonical evolutionary  computations,  of
  $M_{\rm H} \sim 10^{-4} M_*$ at this stellar mass value.

\item The  value of the  thickness of the  H envelope of  our best-fit
  model is in  perfect agreement with the predictions  of the post-LTP
  scenario proposed by Althaus et  al. (2005b) for the formation of DA
  white dwarfs with thin H envelopes.

\item  The luminosity of  our asteroseismological  model allows  us to
  infer a seismological parallax of G117$-$B15A, that is substantially
  larger than  its trigonometric parallax. In  agreement with previous
  works,  we  argue that  the  trigonometric  parallax uncertainty  is
  larger and the seismological derivation of the parallax is robust.

\end{itemize}

As for the complete sample of 44 ZZ Ceti stars, our main results are:

\begin{itemize}

\item We determined the spectroscopic masses of the 44 stars analysed 
using our DA white dwarf evolutionary tracks.

\item  The mean  value  of the  asteroseismological  mass is  $\langle
  M_*\rangle_{\rm seis} = 0.636  \pm 0.019 M_{\odot}$, slightly higher
  than our mean spectroscopic  mass, of $\langle M_*\rangle_{\rm spec}
  =  0.630  \pm  0.028  M_{\odot}$. Given  the  completely  different
  approaches  employed to  derive both  values, the  agreement  can be
  considered as excellent. 

\item Our derived value for $\langle M_*\rangle_{\rm seis}$ is in line
  with  the mean  mass  of DA  white  dwarfs inferred  by Tremblay  et
  al. (2011), $\langle M_*\rangle_{\rm  DA} = 0.613 M_{\odot}$, and in
  good agreement  with the value  derived by Falcon et  al. (2010),
  $\langle M_*\rangle_{\rm DA} = 0.647^{+0.013}_{-0.014} M_{\odot}$.

\item There  exists a  range of  thicknesses of the  H envelope  in the
  studied ZZ Ceti stars, in  qualitative agreement with the results 
  of Castanheira
  \&  Kepler (2009).  Our  distribution of  H envelope  thicknesses is
  characterised  by  a strong  peak  at   thick envelopes  [$\log
  (M_{\rm  H}/M_*)\sim -4.5$]  and  another much less pronounced 
   peak  at very  thin envelopes [$\log (M_{\rm H}/M_*)\sim -9.5$], 
   with an evident paucity for  intermediate thicknesses.

\item In most of the analysed DAVs  (34 stars from a total of 44), our
  asteroseismological models have H  envelopes thinner than the values
  predicted    by   standard    evolutionary    computations   
  for a given stellar mass.  However, our envelope masses 
  cluster closer to the canonical envelope masses than those of 
  Castanheira \& Kepler (2009).

\end{itemize}

In closing, we note that  Tremblay \& Bergeron (2008) have studied the
ratio of He-rich to H-rich white dwarfs in terms of $T_{\rm eff}$ from
a model atmosphere analysis of  the infrared photometric data from the
Two Micron  All Sky Survey combined with  available visual magnitudes.
They found that this ratio increases gradually from $\approx 0.25$ for
$15\,000 \gtrsim  T_{\rm eff}  \gtrsim 10\,000$ K  to about  $0.5$ for
$10\,000  \gtrsim T_{\rm  eff}  \gtrsim 8\,000$  K  due to  convective
mixing when the bottom of the H convection zone reaches the underlying
convective He envelope. These authors  conclude that about $15\%$ of the
DA  white dwarfs should  have H  envelopes with  $\log(M_{\rm H}/M_*)$
between $-10$  and $-8$.  The asteroseismological  results reported in
this work point  to the existence of large fraction  of DAV stars with
H envelopes thinner than canonical values.   
In  particular,  5 ZZ  Ceti  stars analyzed  have
$10^{-10} \lesssim  M_{\rm H}/M_* \lesssim  10^{-8}$, which represents
the $11 \%$ of the sample of the studied DAV stars. This fraction of
stars with very thin H envelopes is compatible with the results
of Tremblay \& Bergeron (2008).

In  a detailed asteroseismological  analysis of  an ensemble  of 
ZZ Ceti stars,
Castanheira \& Kepler  (2008, 2009) have found that  the H envelope of
these  stars could be within  the range $3  \times 10^{-10} \lesssim
M_{\rm  H}/M_* \lesssim 10^{-4}$,  with an  average value  of $\langle
M_{\rm  H}/M_* \rangle=  5  \times 10^{-7}$.   In  many respects,  the
results  of the  present study  are  in excellent  agreement with  the
predictions of Castanheira \& Kepler (2008, 2009).  Our different mean
value  for the  H layer  mass,  $\langle M_{\rm  H}/M_* \rangle=  2.71
\times 10^{-5}$,  which is  about 50 times  larger than that  found by
those authors, could be due to  the fact that our studies are based on
completely  independent  sets  of  DA white  dwarf  models,  different
pulsational codes, and different samples of stars.

All these results reinforce the idea that a non-negligible fraction of
DA  white  dwarfs with  thin  H  envelopes  could exist,  rendering  as
a plausible one the scenario proposed  by Althaus et al. (2005b) for the
formation of DA  white dwarfs with $M_{\rm H}$  smaller than predicted
by the standard theory. Hopefully, new asteroseismological analysis on
a larger  number of DAV  stars, including the  ZZ Ceti stars  from the
SDSS, based on fully evolutionary DA white dwarf models with realistic
chemical profiles  like the ones employed  in this work,  will help to
place this idea on a firmer basis.

\section*{Acknowledgments}

Part of  this work  was supported by  AGENCIA through the  Programa de
Modernizaci\'on   Tecnol\'ogica  BID   1728/OC-AR, by   the  PIP
112-200801-00940  grant from CONICET, and CAPES/MINCyT.  
This research  has made  use of NASA's Astrophysics Data System.

\label{lastpage}

\end{document}